\documentclass[aps, prfluids, showpacs, showkeys, notitlepage, amsmath, amssymb, floatfix, longbibliography, 12pt, tightenlines]{revtex4-2}
\usepackage[export]{adjustbox}
\usepackage{svg}
\usepackage{float}
\usepackage{amsmath}
\usepackage{amssymb}
\usepackage{multirow}
\usepackage[font=small, labelfont=bf, justification=justified]{caption}
\usepackage{subfig}
\usepackage{graphicx}
\usepackage{dcolumn}
\usepackage{bm}
\usepackage{hyperref}
\DeclareMathAlphabet{\altmathcal}{OMS}{cmsy}{m}{n}

\begin{document}

\title{Dynamical regimes and clustering of small neutrally buoyant inertial particles in stably stratified turbulence}
\author{Christian Reartes and Pablo D. Mininni}
\affiliation{Universidad de Buenos Aires, Facultad de Ciencias Exactas y Naturales, Departamento de Física, Ciudad Universitaria, 1428 Buenos Aires, Argentina,}
\affiliation{CONICET - Universidad de Buenos Aires, Instituto de F\'{\i}sica del Plasma (INFIP), Ciudad Universitaria, 1428 Buenos Aires, Argentina.}

\begin{abstract}
Inertial particles in stably stratified flows play a fundamental role in geophysics, from the dynamics of nutrients in the ocean to the dispersion of pollutants in the atmosphere. We consider the Maxey-Riley equation for small neutrally buoyant inertial particles in the Boussinesq approximation, and discuss its limits of validity. We show that particles behave as forced damped oscillators, with different regimes depending on the particles Stokes number and the fluid Brunt-V\"ais\"al\"a frequency. Using direct numerical simulations we study the particles dynamics and we show that small neutrally buoyant particles in these flows tend to cluster in regions of low local vorticity. The particles, albeit small, behave fundamentally differently than tracers.
\end{abstract}

\maketitle

\section{Introduction}

Dispersion of inertial particles by turbulent flows plays a fundamental role in many geophysical systems, from cloud formation and the dispersion of pollutants in the atmosphere, to the dynamics of plankton in the ocean \cite{wyngaard_1992, dasaro_2000, watanabe_2017, amir_2017}. In spite of their interest, the dynamics of particles in these systems is poorly understood. We know the equations of motion of small particles (i.e., such that the Reynolds number at the particle scale is much smaller than unity \cite{1}), and in recent years experiments \cite{Obligado_2015} and particle-resolved simulations \cite{Tavanashad_2021} have provided valuable insights into particles' dynamics in other regimes. However, particle transport problems in geophysics necessarily require reduced models that simplify the physics, as even in the cases in which a simulation can be done resolving a broad range of scales, an ensemble of runs to get statistical information becomes rapidly unfeasible. This has resulted, e.g., for stratified flows as in the oceans and the atmosphere, in the modeling of inertial particles simply as Lagrangian trancers \cite{Wagner_2019} or using simplified models \cite{Palmer_2019, Beron_2019}.

Modeling geophysical flows pose other challenges, even without particles. Stably stratified turbulence is anisotropic, and as a result it is fundamentally different from homogeneous isotropic turbulence (HIT) \cite{lindborg_2008, marino_2014, Portwood_2019}. In these flows stratification reduces the vertical velocity, confining the flow into a quasi-horizontal layered motion,  also generating vertically sheared horizontal winds (VSHWs) with strong vertical variability \cite{smith_2002}. The stratification results in a restoring force, allowing for the excitation of waves that can coexist with the turbulence. The spectral scaling of stably stratified turbulence is also different than in HIT, with a rich behavior depending on the scale considered, and on many dimensionless parameters. In broad terms, stably stratified turbulence displays an anisotropic subrange with a direct energy cascade between the buoyancy and Ozmidov scales \cite{waite_2011, maffioli_2017}. Studies also indicate that larger scale quasi-horizontal motions can be a continuous source of small scale turbulence as long as the local Reynolds number does not drop below a threshold \cite{riley_2003}.

Many recent studies have considered stably stratified flows from a Lagrangian perspective (i.e., by considering tracers, or particles with inertia, that are transported by such flows). As an example, the Lagrangian transport of tracers in stably stratified turbulence was studied in \cite{van_aartrijk_2008, Sujo_2018}. However, the general equations of motion for inertial particles submerged in turbulent flows are not clear. For very small particles the Maxey-Riley approximation provides a set of equations for their dynamics \cite{1}. As particles become larger, Basset-Boussinesq and Faxen corrections become relevant, but for even larger particles such perturbative expansion breaks down. The case of stably stratified turbulence is simpler than HIT in some way: most particles and aerosols are much smaller than the dissipation scale, and thus the Maxey-Riley approximation should hold (except for, e.g., large rain droplets in clouds, or snowflakes). But even in this regime, the derivation of the equations requires certain approximations and depend on the form of the equations for the fluid \cite{van_aartrijk_2010, Sozza2016}. 

A fundamental feature of particles in homogeneous and isotropic turbulent flows is their preferential concentration. Turbulence sometimes separates the particles instead of mixing them. The detailed mechanisms by which turbulence affects particle motions are still unclear. In the homogeneous and isotropic case, and for the average concentration, the main mechanisms behind heavy particles clustering are centrifugal expulsion \cite{1} and the sweep-stick mechanism \cite{sweep}. Evidence of preferential concentration of heavy particles in laboratory experiments and numerical simulations was reported, e.g., in \cite{obligado}. Multiscale flow effects may be also relevant \cite{Bragg_2015, Tom_2019}, and the role of other effects in preferential concentration, such as finite particle radius or the effect of large-scale flows, are still unclear \cite{n1, n2, sofi}. In the particular case of stratified turbulence it has been shown that inertial particles also cluster for a wide range of parameters \cite{Sozza2016}. Vertical confinement caused by density stratification produces strong fractal clustering at isopycnic surfaces. Clustering was found to depend on a single parameter, the combination of the Stokes time $\tau_p$ of the particles and the Brunt-Väisälä frequency of the flow. In the limit of small $\tau_p$ (i.e., small inertia), clustering was found to increase monotonically with $\tau_p$ \cite{Sozza2016}.

In this work we present a study considering the Maxey-Riley model for small inertial particles \cite{1}, from which we derive an equation for the dynamics of inertial particles in stably stratified flows. We then perform direct numerical simulations of the Boussinesq equations for the fluid, together with the Maxey-Riley equation for one million particles. We derive a simple model for the particles vertical displacement, and compare the model with the simulations to show that particles behave as forced damped oscillators with different regimes depending on the Stokes and Froude numbers. We also characterize the dependence of the stratification-induced vertical confinement of the particles on these two parameters. Finally, we study the formation of clusters using Voronoi tessellation, and show that particles in stably stratified flows tend to accumulate in regions with low vorticity, at least for the range of parameters considered in the present study.

\section{Equations of motion \label{sec:theory}}

In this work we solve numerically the incompressible Boussinesq equations for the velocity $\bf u$ and for mass density fluctuations $\rho'$,
\begin{equation}
\partial_t {\bf u} + {\bf u} \cdot \boldsymbol{\nabla} {\bf u} = - \boldsymbol {\nabla} \left(p/\rho_0\right) - \left(g/\rho_0 \right)\rho' \hat{z}+ \nu \nabla^2 {\bf u} + {\bf f},
\end{equation}
\begin{equation}
\partial_t \rho'+ {\bf u} \cdot \boldsymbol{\nabla} \rho' = \left(\rho_0 N^2/g\right) {\bf u}\cdot \hat{z}+ \kappa \nabla^2 {\bf \rho'},
\end{equation}
\begin{equation}
\nabla \cdot {\bf u} = 0,
\end{equation}
where $p$ is the correction to the hydrostatic pressure, $\nu$ is the kinematic viscosity, ${\bf f}$ is an external mechanical forcing, $N$ is the Brunt-V\"{a}is\"{a}l\"{a} frequency (which in this approximation sets the stratification), and $\kappa$ is the diffusivity. In terms of the background density gradient, the Brunt-V\"{a}is\"{a}l\"{a} frequency is $N^2 = -(g/\rho_0) (d\bar \rho /dz)$, with $d \bar \rho /dz$ the imposed (linear) background stratification, and $\rho_0$ the mean fluid density. We write scaled density fluctuations $\zeta$ in units of velocity by defining $\zeta = g\rho'/(\rho_0 N)$.  All quantities are then made dimensionless using a characteristic length $L_{0}$ and a characteristic velocity $U_{0}$ in the domain, resulting in
\begin{equation}
\partial_t {\bf u} + {\bf u} \cdot \boldsymbol{\nabla} {\bf u} = - \boldsymbol {\nabla} \left(p/\rho_0\right) - N\zeta \hat{z}+ \nu \nabla^2 {\bf u} + {\bf f},
\label{bou1}
\end{equation}
\begin{equation}
\partial_t \zeta + {\bf u} \cdot \boldsymbol{\nabla} \zeta = N {\bf u}\cdot \hat{z}+ \kappa \nabla^2 {\bf \zeta}.
\label{bou2}
\end{equation}

Inertial particles are modeled using the Maxey-Riley model, but we consider an approximation consistent with those made to obtain the Boussinesq equations, in addition to assuming that the typical length over which the velocity field changes appreciably is much larger than the particle radius $a$. Under the latter hypothesis the Fax\'en terms are negligible. Under the Boussinesq approximation for a stratified flow, Eqs.~(\ref{bou1}) and (\ref{bou2}) are obtained from the Navier-Stokes equations after neglecting all density fluctuations except for those in the buoyancy force. Thus, for the dynamics of the particles we also consider the density and the mass of the fluid displaced by the particles in terms of their mean values, respectively $\rho_f  \approx \overline{\rho}_f = \rho_0$ and $m_f \approx \overline{m}_f = \rho_0 V_p$ (where $V_p$ is the volume of the particles), except in the gravity term. In that term we consider the entire fluid density dependence, $\rho_f = \rho_0 + \partial\bar{\rho}/\partial z(z-z_0)+\rho'$, for a linear background density profile. As the flow is stably stratified, $\partial\bar{\rho}/\partial z < 0$. Under these approximations the equation for the particles results in
\begin{equation*}
\Dot{\bf v}\left(1 + \frac{1}{2} \frac{\bar{m}_f}{m_p} \right) = \frac{6\pi a \bar{\rho}_f \nu}{m_p} \left[ {\bf u}({\bf x},t) - {\bf v}(t) \right] +\frac{3}{2}\frac{\bar{m}_f}{m_p} \frac{\textrm{D}}{\textrm{D}t} {\bf u}({\bf x},t)
\label{eqn:fin}
\end{equation*}
\begin{equation}
 - g\left[1 - \frac{1}{\rho_p}\left(\rho_0 + \frac{\partial\bar{\rho}}{\partial z}(z-z_0)+\rho' \right) \right] \hat{z} + \frac{6 \pi {a}^{2} \bar{\rho_f} \nu}{m_p}\int_{0}^{t} \frac{d}{d\tau} [{\bf u}({\bf x},\tau) - {\bf v}(\tau)] \frac{d\tau}{\sqrt{\pi \nu (t-\tau})},
\label{eqn:mr}
\end{equation}
where ${\bf x}$ is the particle position, ${\bf v}$ is the particle velocity, ${\bf u}({\bf x},t)$ is the fluid velocity at the particle position, $D/Dt$ is the Lagrangian derivative, $d/dt$ is the time derivative following the particle trajectory, and $\rho_p$ is the particle mass density (particles are assumed to be spherical). For a fluid at rest, note particles will be at equilibrium (i.e., neutrally buoyant) when $1- \rho_f/\rho_p = 0$, and that there is some freedom on how $\rho_0$ and $z_0$ are chosen. In particular, without loss of generality we can choose $\rho_p = \rho_0$, such that particles are neutrally buoyant at $z=z_0$ in the absence of density fluctuations.

Multiplying and dividing the buoyancy term in Eq.~(\ref{eqn:mr}) by $\rho_0$ we have,
\begin{equation*}
\Dot{\bf v}\left(1 + \frac{1}{2} \frac{\bar{m}_f}{m_p} \right) = \frac{6\pi a \bar{\rho}_f \nu}{m_p} \left[ {\bf u}({\bf x},t) - {\bf v}(t) \right] +\frac{3}{2}\frac{\bar{m}_f}{m_p} \frac{\textrm{D}}{\textrm{D}t} {\bf u}({\bf x},t)
\label{eqn:fin}
\end{equation*}
\begin{equation}
 - \frac{\rho_0}{\rho_p}\left[\frac{g}{\rho_0} \frac{\partial\bar{\rho}}{\partial z}(z-z_0)+g\frac{\rho'}{\rho_0} \right] \hat{z} + \frac{6 \pi {a}^{2} \bar{\rho_f} \nu}{m_p}\int_{0}^{t} \frac{d}{d\tau} [{\bf u}({\bf x},\tau) - {\bf v}(\tau)] \frac{d\tau}{\sqrt{\pi \nu (t-\tau})},
\label{eqn:fin}
\end{equation}
where the first term inside the brackets in the buoyancy is $ -N^2 $, while the second term is $N \zeta$. Reordering the terms in the equation and using dimensionless units we finally obtain,
\begin{equation}
\Dot{\bf v} = \frac{1}{\tau_p} \left[ {\bf u}({\bf x},t) - {\bf v}(t) \right] - \frac{2}{3}N\left[N(z-z_0) - \zeta \right] \hat{z} + \frac{\textrm{D}}{\textrm{D}t} {\bf u}({\bf x},t) +\sqrt{\frac{3}{\pi \tau_p}}\int_{0}^{t} d\tau \frac{\frac{d}{d\tau} [{\bf u}({\bf x},\tau) - {\bf v}(\tau)]}{\sqrt{t-\tau}} ,
\label{ec.parts}
\end{equation}
where the particle relaxation time is $\tau_p = (m_p + \bar{m}_f/2)/(6 \pi a \bar{\rho}_f \nu)$. For a spherical particle $\tau_p = a^2/(3 \nu)$, with $\gamma = \bar{m}_f/m_p = 1$. We define the Stokes number as $\textrm{St} = \tau_p/\tau_\eta$, where $\tau_\eta = (\nu/\varepsilon)^{1/2}$ is the Kolmogorov time scale and $\varepsilon$ is the fluid kinetic energy dissipation rate. Note that any other choice for $\gamma = \bar{m}_f/m_p$ is equivalent to changing the reference value $\rho_0$, and results in the particles being neutrally buoyant at a different height (or equivalently, it results in a redefinition of $z_0$).

\section{Numerical set up \label{sec:methods}}

Besides the Stokes number that characterizes the particles, Eqs.~(\ref{bou1}) and (\ref{bou2}) have two controlling parameters for the fluid, the Reynolds and Froude numbers,
\begin{equation}
 \textrm{Re} = \frac{LU}{\nu}, \,\,\,\,\,\, 
 \textrm{Fr} = \frac{U}{LN},
 \label{eq:Re_Fr}
\end{equation}
where $L$ and $U$ are respectively the characteristic Eulerian integral length and the r.m.s.~flow velocity. Using these parameters we can also define the buoyancy Reynolds number
\begin{equation}
 \textrm{Rb}  =   \textrm{Re} \, \textrm{Fr}^{2},
\end{equation}
which provides an estimation of how turbulent the flow is at the buoyancy scale $L_{b}=U/N$, and plays an important role characterizing the flow dynamics. For $\textrm{Rb} \gg 1$ strong stratified turbulence can develop, while for $\textrm{Rb} \ll 1$ turbulent motions are strongly damped by viscosity. Geophysical flows typically have large $\textrm{Rb}$; observations in the ocean thermocline yield $\textrm{Rb} \approx 10^2$ to $10^3$ \cite{moum_1996}. Considering numerical limitations, here we study flows with $\textrm{Rb}>10$.  The Ozmidov scale, $L_{oz}=2\pi/k_{oz}$ (with $k_{oz}=\sqrt{N^3/\varepsilon}$) also plays an important role in the dynamics, as for scales sufficiently small compared with $L_{oz}$ the flow is expected to recover isotropy. For $\textrm{Rb}>1$, $L_{oz}$ is larger than the Kolmogorov dissipation scale $\eta$, and quasi-isotropic turbulent transport can be expected to take place at small scales.

Setting these parameters and choosing the forcing prescribes the numerical simulations. For the forcing we use Taylor-Green forcing \cite{clark_di_leoni_2015}, which is a two-velocity components forcing that generates pairs of large-scale counter-rotating eddies perpendicular to the stratification, with a shear layer in between. Its expression is given by
\begin{equation}
    {\bf f} = f_{0} \left[ \sin(k_{f}x)\cos(k_{f}y)\cos(k_{f}z) \hat{\bf x} - \cos(k_{f}x)\sin(k_{f}y)\cos(k_{f}z) \hat{\bf y} \right],
\end{equation}
where $f_0$ is the amplitude of the forcing, $k_f=1/L_0$ is the forcing wave number, and $L_0$ a unit length. This flow has been used before to study stratified turbulence \cite{riley_2003, Sujo_2018}. In the stratified case it generates a large-scale circulation with VSHWs (i.e., with a non-zero mean horizontal velocity) only in the shear layer between the large-scale Taylor-Green vortices (see \cite{sm} for a movie of the development of the non-zero mean horizontal wind in this layer).

The Boussinesq fluid equations, Eqs.~(\ref{bou1}) and (\ref{bou2}), were solved in a triply periodic domain using a parallelized and fully dealiased pseudo-spectral method, and a second-order Runge-Kutta scheme for time integration \cite{mininni_hybrid_2011}. The equation for the particles,  Eq.~(\ref{ec.parts}), was solved using third-order spline interpolation to estimate the forces at the particles positions, and with a second-order Runge-Kutta method for the time evolution \cite{Yeung_1988}.

\begin{table*}
\caption{\label{tablaga} Relevant parameters of the fluid simulations. $NT_0$ is the Brunt-Väisälä frequency in units of $T_0^{-1}=U_0/L_0$, $\textrm{Fr}$ is the Froude number, $\textrm{Re}$ is the Reynolds number, $\textrm{Rb}$ is the buoyancy Reynolds number, $L$ is the flow integral scale, $\eta$ is the Kolmogorov scale, $L_b$ is buoyancy length, and $L_{Oz}$ is the Ozmidov length scale. All lengths are in units of the unit length $L_0$.}
\begin{ruledtabular}
\begin{tabular}{ccccccccc}
 Run  & $N T_0$ & $\textrm{Fr}$ & $\textrm{Re}$ & $\textrm{Rb}$ & $L/L_0$ & $\eta/L_0$ & $L_b/L_0$ & $L_{Oz}/L_0$ \\ \hline
$N04$  & 4  & 0.20 & 1900 & 76 & 1.27 & 0.0072 & 0.25 & 0.30 \\ 
$N08$  & 8  & 0.12 & 1700 & 24 & 1.07 & 0.0072 & 0.13 & 0.10 \\
$N12$  & 12 & 0.09 & 1600 & 13 & 1.00 & 0.0072 & 0.09 & 0.06
\label{tab_fluid}
\end{tabular}
\end{ruledtabular}
\end{table*}

\begin{table*}
\caption{\label{tablaga} Relevant parameters of the particles in all simulations. $\textrm{St}$ is the Stokes number, $\tau_p/T_0$ is the Stokes time in units of $T_0$, $a/\eta$ is the particle radius in units of the Kolmogorov scale, and $\textrm{Re}_p$ are the respective particles Reynolds numbers for all the Brunt-Väisälä frequencies.}
\begin{ruledtabular}
\begin{tabular}{ccccccc}
\multirow{2}{*}{Label} & \multirow{2}{*}{$\textrm{St}$} & \multirow{2}{*}{$\tau_p/T_0$} & \multirow{2}{*}{$a_p/\eta$} & \multicolumn{3}{c}{$\textrm{Re}_p$}\\
\cline{5-7}
   &   &   &   & $NT_0=4$ & $NT_0=8$ & $NT_0=12$\\
\hline
$St03$  & 0.3 & 0.024 & 0.96 & 0.2 & 0.2 & 0.1\\ 
$St1$   & 1   & 0.076 & 1.67 & 0.7 & 0.5 & 0.2 \\ 
$St3$   & 3   & 0.235 & 3.03 & 4.2 & 2.7 & 1.6\\
$St6$   & 6   & 0.470 & 4.28 & - & 7.6 & -
\label{tab_parts}
\end{tabular}
\end{ruledtabular}
\end{table*}

We performed several direct numerical simulations of the Boussinesq equations with different Froude numbers, using a spatial resolution of $N_x = N_y = 768$ and $N_z = 192$ grid points, in a triple periodic domain of length $L_x=L_y=2\pi L_0$ in the horizontal directions and $L_z=H=\pi L_0/4$ in the vertical direction. Three different Brun-V\"ais\"al\"a frequences are considered (times are measured in units of a unit turnover time $T_0 = L_0/U_0$, with $U_0$ a unit velocity, see Table \ref{tab_fluid} for all the relevant fluid parameters). All simulations have a Prandtl number $\textrm{Pr} = \nu/\kappa = 1$, with the kinematic viscosity chosen so that the Kolmogorov scale $\eta = (\nu^3/\varepsilon)^{1/4} \approx 0.0072 L_0$ is well resolved, where the kinetic energy dissipation rate is computed as $\varepsilon = \nu \left< |\boldsymbol{\omega}|^2\right>$ and $\boldsymbol{\omega} = \boldsymbol{\nabla} \times {\bf u}$ is the vorticity. This results in $\kappa \eta \approx 1.9$, where $\kappa = N_x/(3L_0)$ is the maximum resolved wave number, corresponding to spatially well resolved simulations \cite{Donzis_2010, Wan_2010}.

Once the flows in these simulations have reached a turbulent steady state, we randomly distributed particles in a horizontal strip of width $H/5$ centered around $z_0 = H/2$ (i.e., at the height of the shear layer of the Taylor-Green flow), with initial velocities equal to the fluid velocity at the center of each particle. The dynamics of the particles neglected the last term in Eq.~(\ref{ec.parts}) (i.e., the Basset-Boussinesq history term). However, that term was computed {\it a posteriori} to estimate its relevance in the dynamics. Particles were one-way coupled, and thus they can be considered as test particles: they do not collide, and their volume fraction is irrelevant for the flow dynamics. Thus, the number of particles should be considered solely as a way to improve the statistics. To each simulation in Table \ref{tab_fluid} we added four sets with $10^6$ particles each, with four different values of $\tau_p$ (or equivalently, different Stokes numbers), resulting in a total of 10 data sets of particles with different $\textrm{Fr}$ and $\textrm{St}$ numbers. Table \ref{tab_parts} lists the relevant parameters of all the simulations with particles, including the particle Reynolds number $\textrm{Re}_p = a|\bf{u}-\bf{v}|/\nu$ in each case. Note that Tables \ref{tab_fluid} and \ref{tab_parts} should be read together, as we can have, e.g., particles with $\textrm{St}=0.3$ in a flow with $N=4/T_0$, or the same particles in a flow with $N=8/T_0$ or $12/T_0$. As a rule, the Basset-Boussinesq force is smaller than the drag force in all simulations except in the cases with $\textrm{St}=3$; for those particles the Basset-Boussinesq history term becomes comparable to the Stokes drag (even though both terms are smaller than buoyancy and added mass forces), and thus studying particles with larger $\textrm{St}$ would require taking this force into account in the dynamics (see, e.g., \cite{van_aartrijk_2010}).

\section{Spectra and particle vertical displacement model}

\begin{figure}
\includegraphics[width=0.46\textwidth]{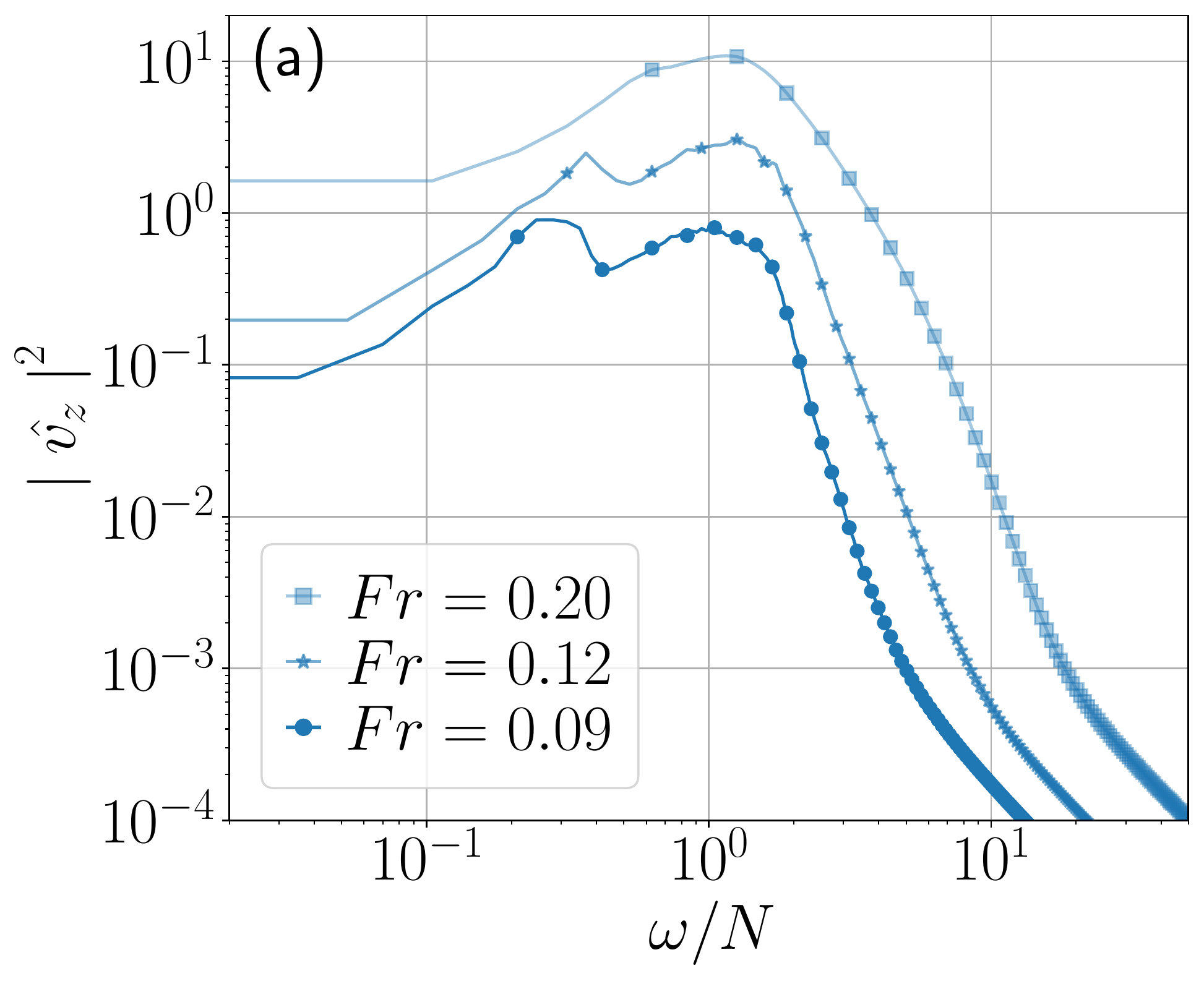}
\includegraphics[width=0.46\textwidth]{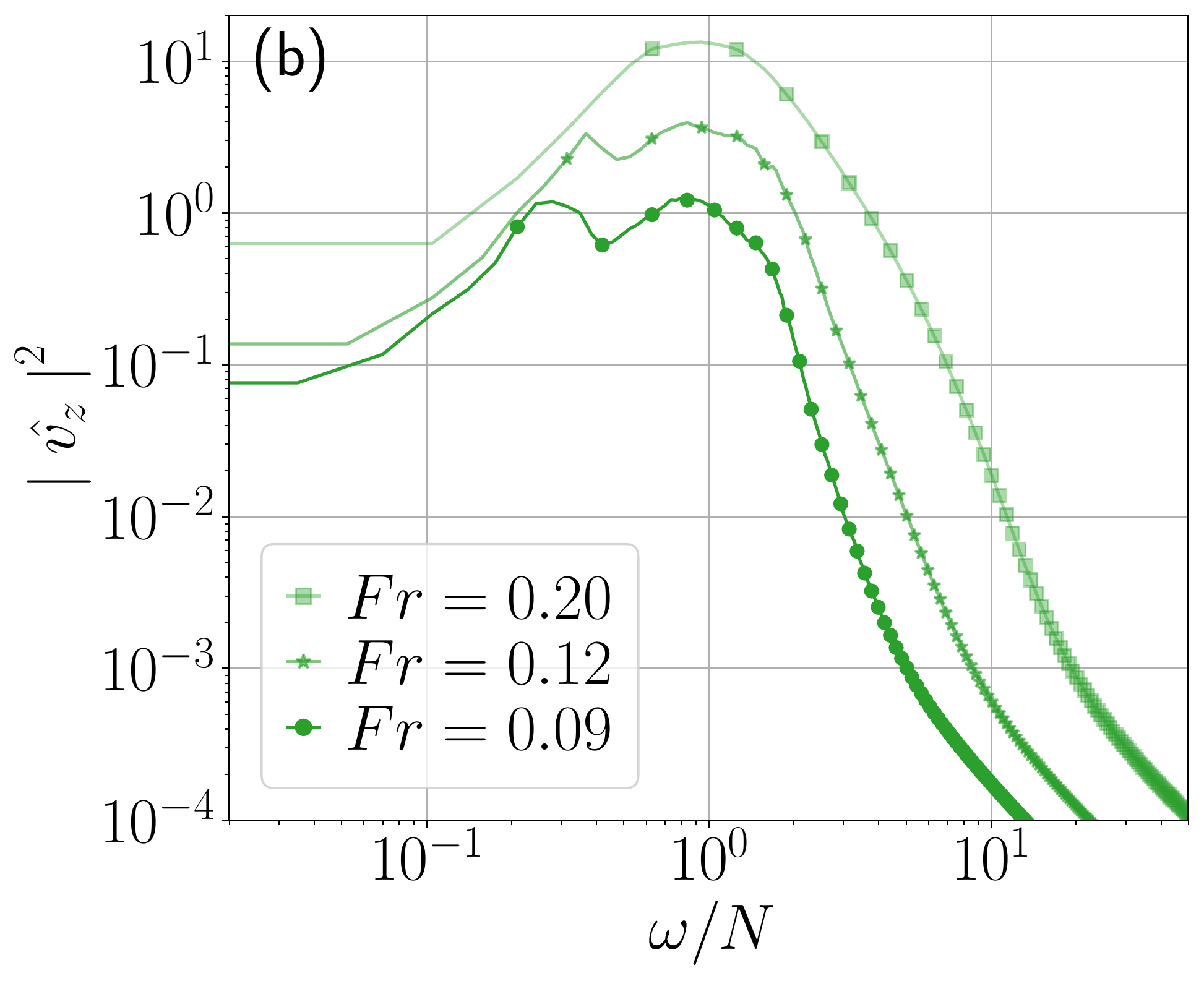}
\includegraphics[width=0.46\textwidth]{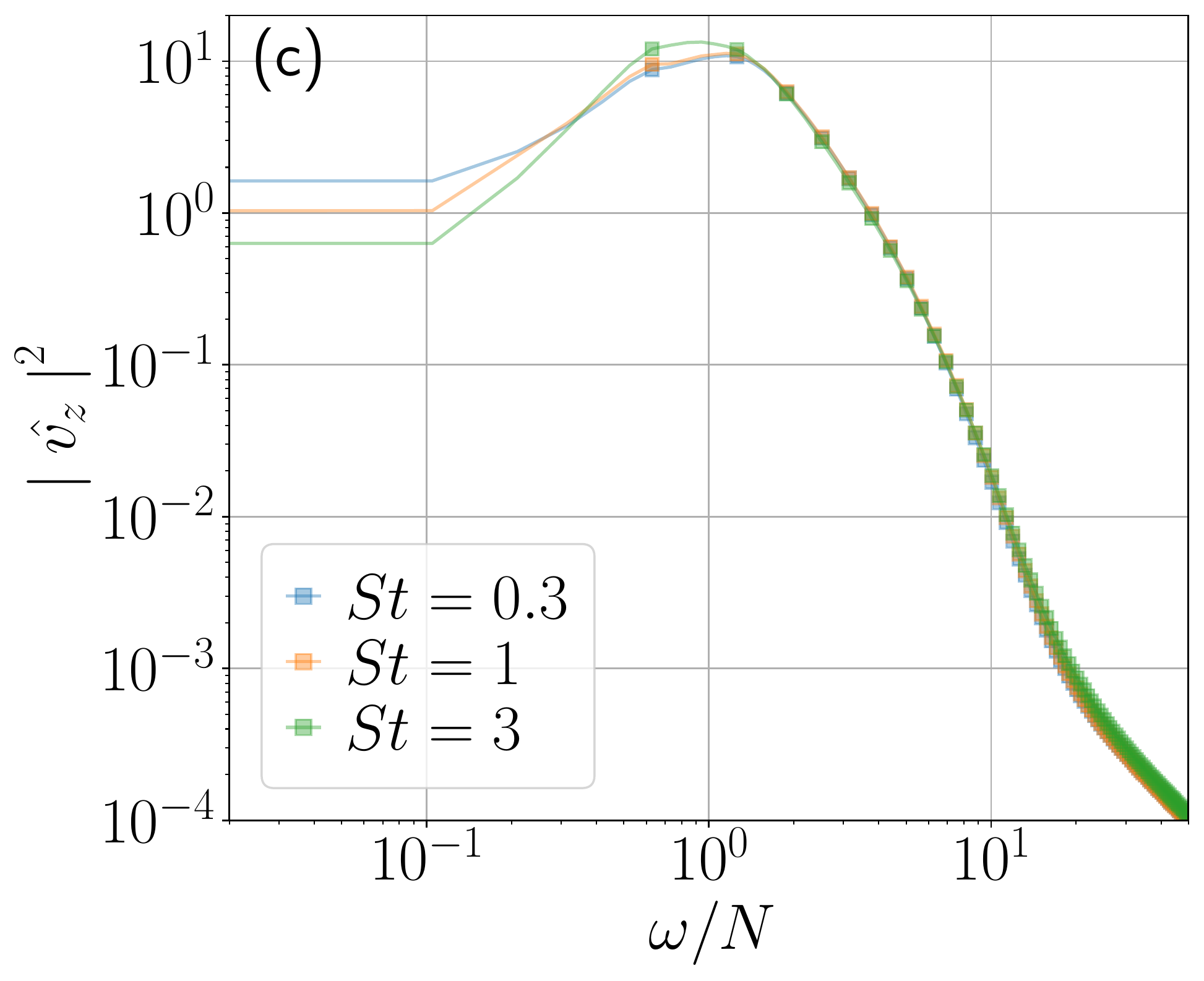}
\includegraphics[width=0.46\textwidth]{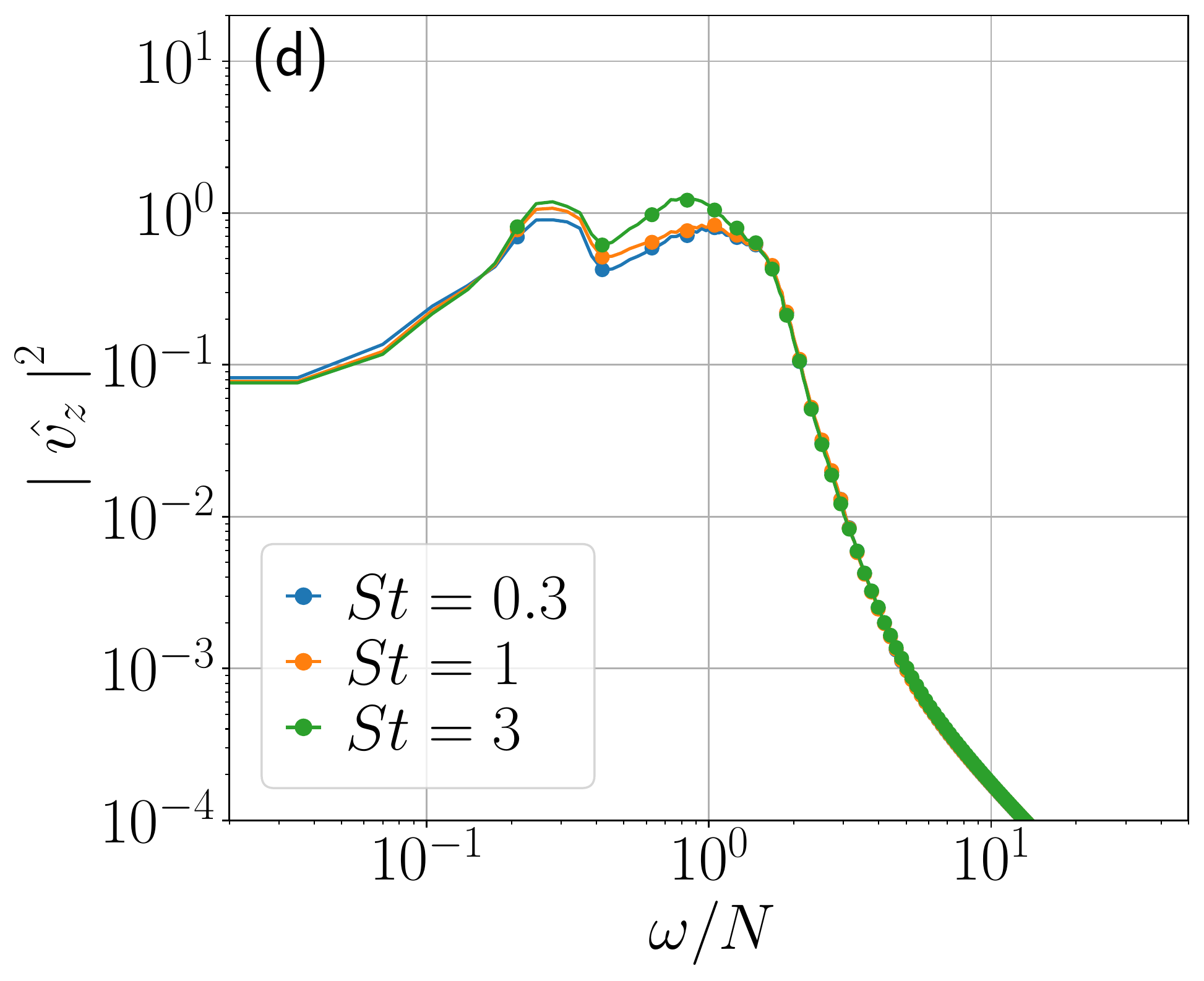}
\caption{Power spectra of the particles' vertical velocity, for different values of the Froude and Stokes numbers. (a) $\textrm{St}=0.3$ and (b) $\textrm{St}=3$, for different $\textrm{Fr}$ (indicated in the insets). Power decreases with increasing stratification. (c) $\textrm{Fr}=0.20$ and (d) $\textrm{Fr}=0.09$, for different $\textrm{St}$ (indicated in the insets). The value of $\textrm{St}$ has a small effect in the amplitude of the main peak.}
\label{spectrum}
\end{figure}

We first study the power spectrum of the particles' vertical velocity. Figure \ref{spectrum} shows this spectrum for different values of the Froude and Stokes numbers; frequencies are normalized by the Brunt-Väisälä frequency of the carrier flow. A peak is always present at $\omega \approx N$, and for small $\textrm{Fr}$ a second peak at lower frequencies is observed. Its position and amplitude depends on $\textrm{Fr}$, while its amplitude depends only weakly on $\textrm{St}$. The peak at $\omega \approx N$ is followed for larger frequencies by a steep spectrum, and decays slowly for smaller frequencies. 

\begin{figure}
\includegraphics[width=0.46\textwidth]{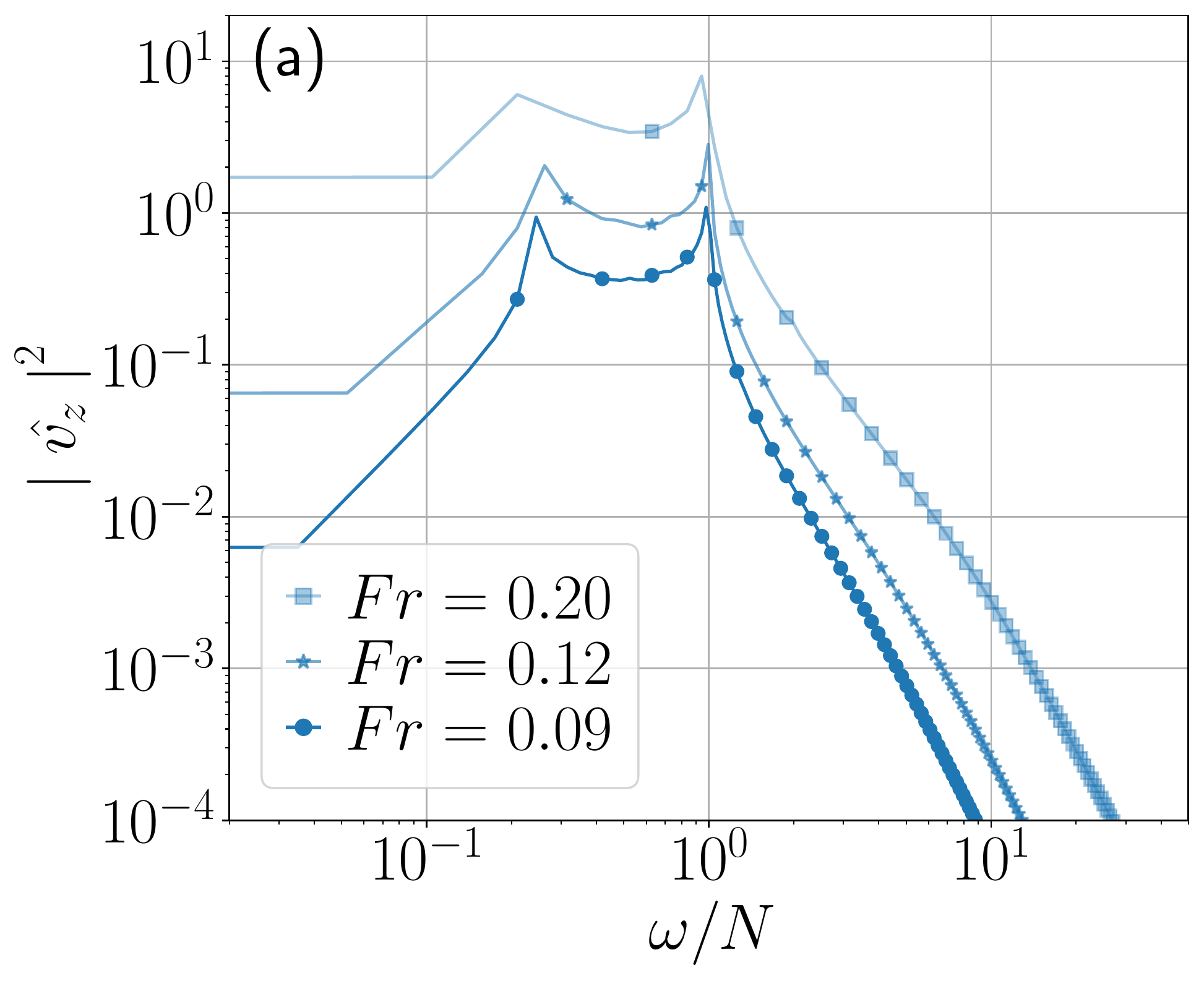}
\includegraphics[width=0.46\textwidth]{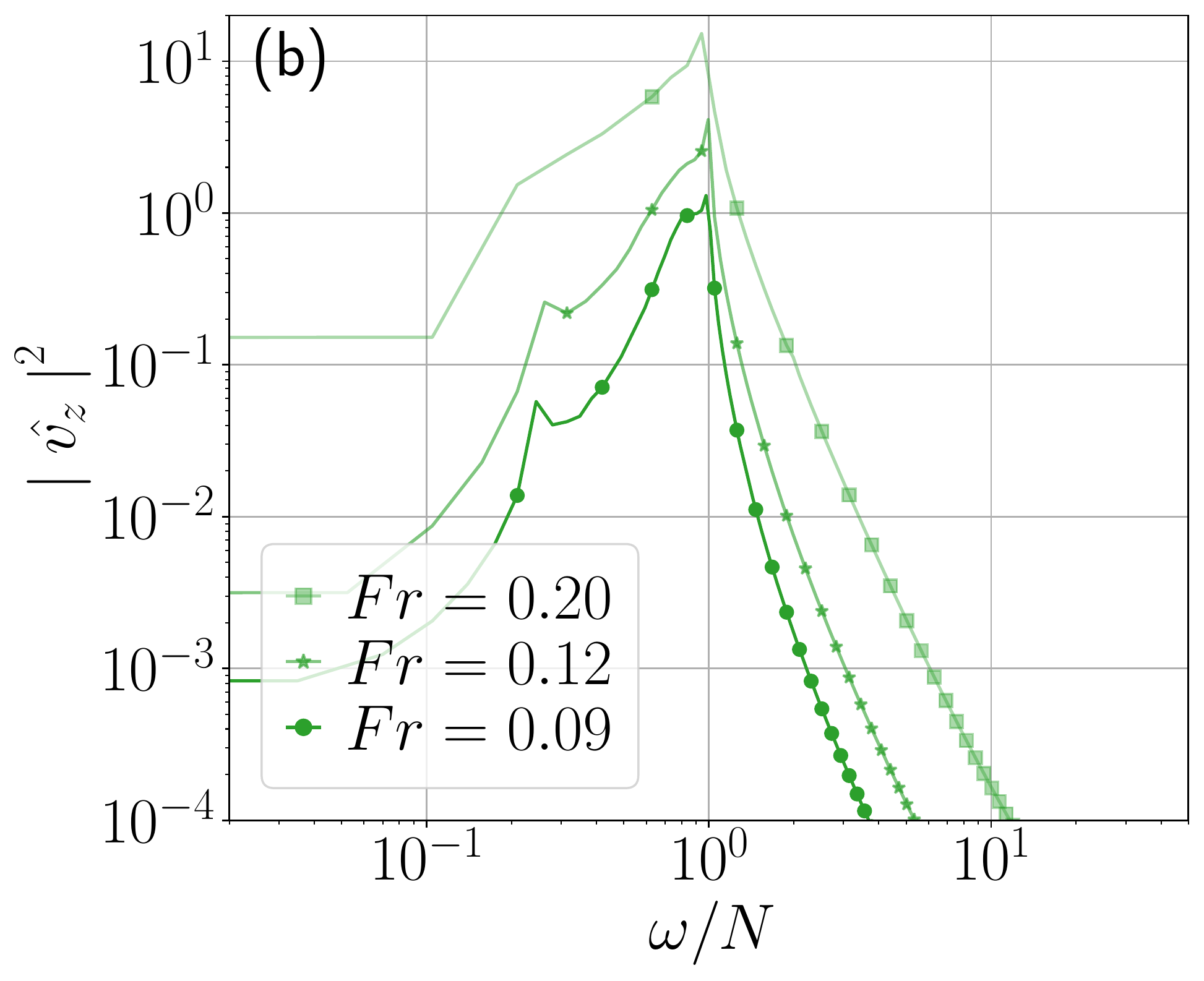}
\includegraphics[width=0.46\textwidth]{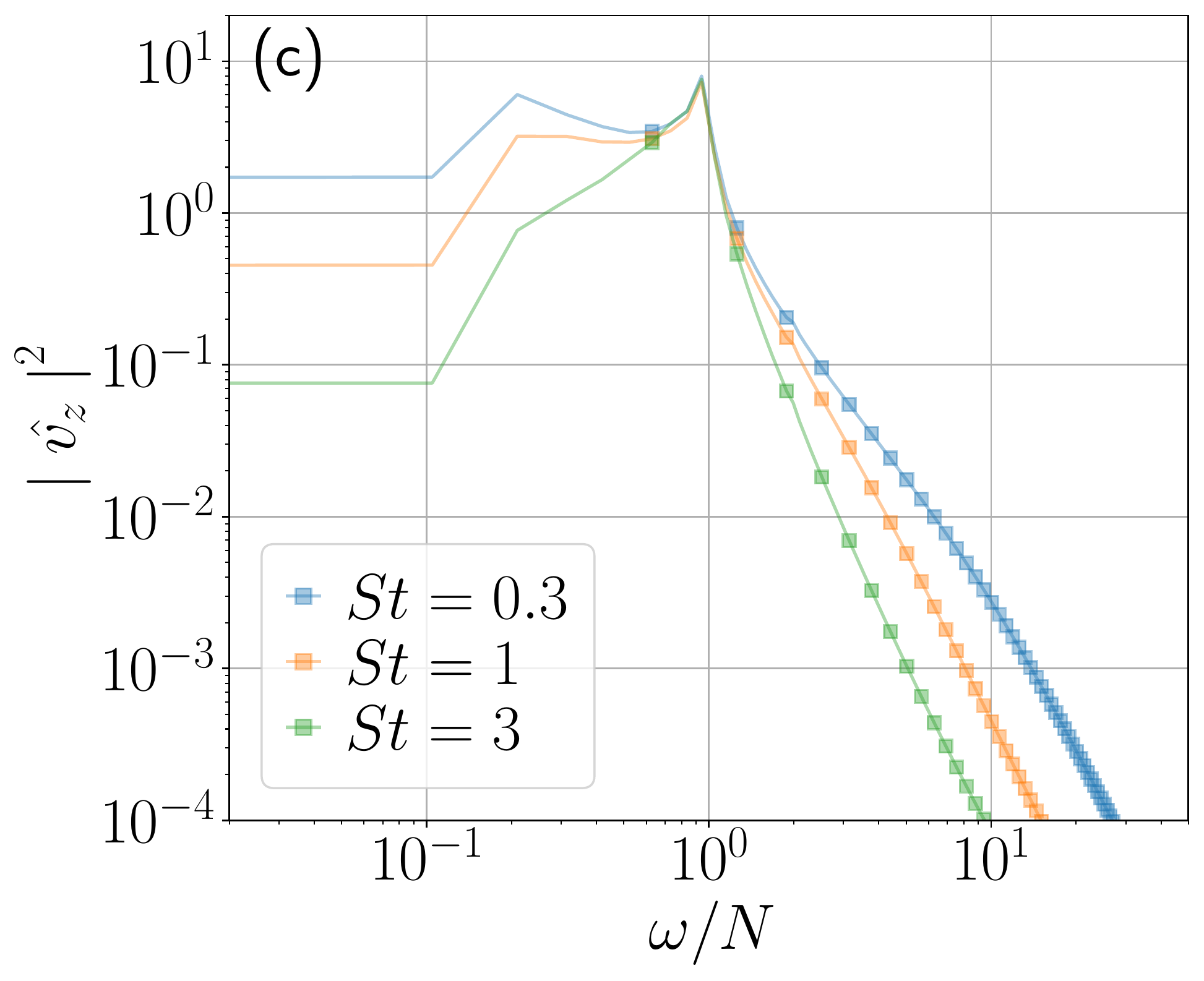}
\includegraphics[width=0.46\textwidth]{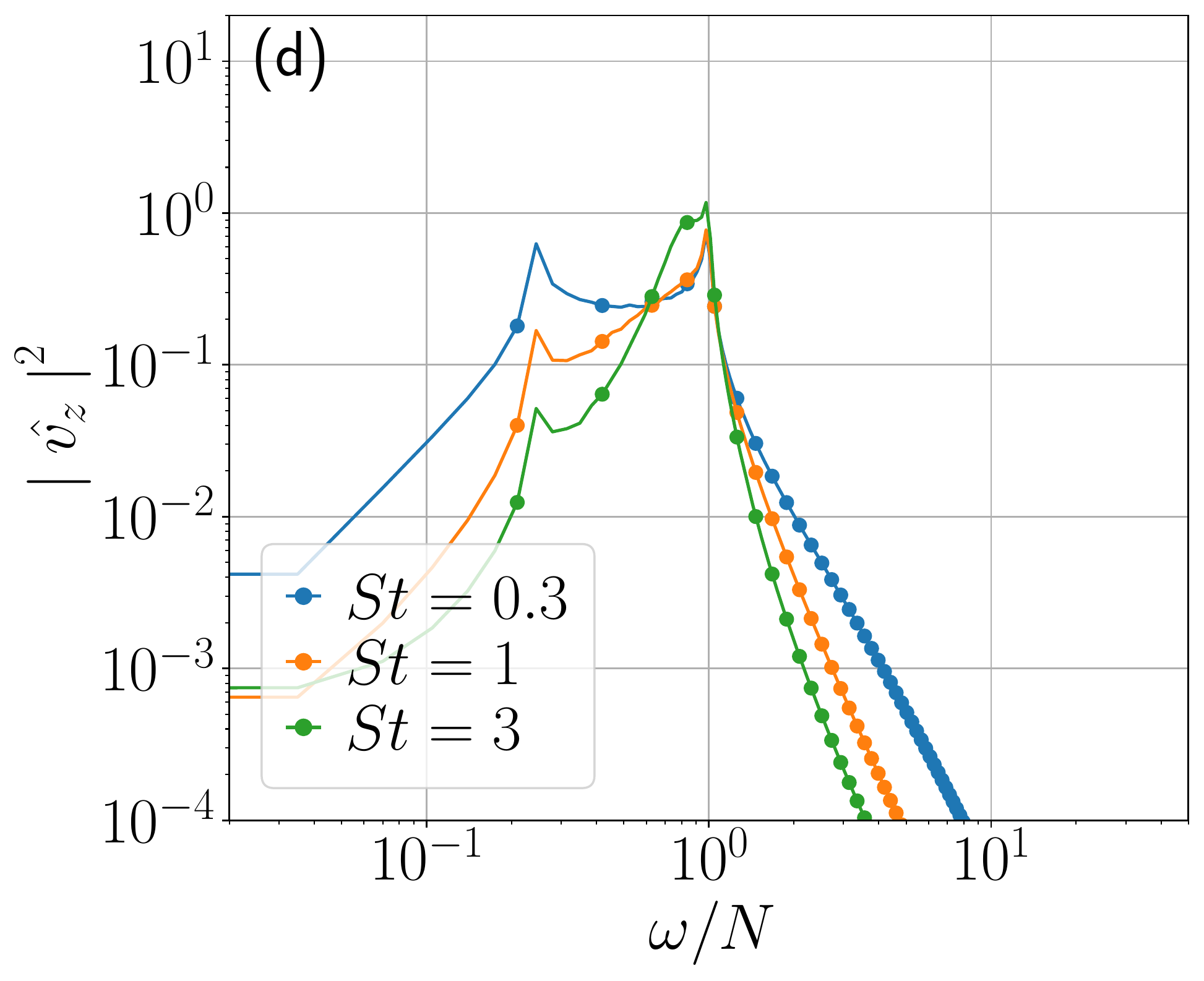}
\caption{Power spectrum of the particles' vertical velocity from the model in Eq.~(\ref{ec.osc}) for different values of $\textrm{Fr}$ and (a) $\textrm{St}=0.3$, and (b) $\textrm{St}=3$. Note the decrease in the power with decreasing $\textrm{Fr}$. Same for different values of $\textrm{St}$ and for (c) $\textrm{Fr} = 0.20$ and (d) $\textrm{Fr} = 0.09$. The value of $\textrm{St}$ changes the frequency of the main peak and the amplitude of the secondary peak.}
\label{model_spectrum}
\end{figure}

The origin of the two peaks in the spectra can be explained by a simple model derived from the equation of motion of the particles. Equation (\ref{ec.parts}) can be rewritten in term of the particles' vertical position $z$ using that $\dot{z} = v_z$ and $\ddot{z} = \dot{v_z}$, resulting in
\begin{equation}
\ddot{z} = \frac{1}{\tau_p} \left[ { u_z}({\bf x},t) - \dot{z} \right] - \frac{2}{3} N\left[N(z-z_0) - \zeta \right] + \frac{\textrm{D}}{\textrm{D}t} { u_z}({\bf x},t),
\label{ec.aprox}
\end{equation}
where the Basset-Boussinesq history force was neglected. Rearranging terms in Eq.~(\ref{ec.aprox}) we arrive at the following expression,
\begin{equation}
\ddot{z} + \frac{1}{\tau_p} \dot{z} + \frac{2}{3}N^2z = \ddot{z}_{wav} + \frac{1}{\tau_p} \dot{z}_{wav} + \frac{2}{3}N^2z_{wav},
\label{ec.osc}
\end{equation}
where we assumed that vertical displacements of fluid elements are caused by internal gravity waves, and thus we defined $z_{wav} = z_0 + \zeta/N$, $\dot{z}_{wav} = u_z({\bf x},t)$, and $\ddot{z}_{wav} = Du_z/Dt$. Equation (\ref{ec.osc}) is the equation of a driven damped oscillator with system frequency $\sqrt{2/3} N$, damping constant $(2 \tau_p)^{-1}$, and forcing $f_{wav} = \ddot{z}_{wav} + \dot{z}_{wav}/\tau_p + 2 N^2 z_{wav}/3$. The pulsation of the damped system is $\Omega^2 = 2N^2/3 - (2\tau_p)^{-2}$. For particles with small inertia this results in an over-damped system (i.e., $\Omega^2<0$) and when perturbed, particles slowly decay to the equilibrium position following fluid elements. Particles with large inertia result instead in weak damping ($\Omega^2>0$), and perturbed particles oscillate around the equilibrium as they decay, only weakly following the fluid elements. Indeed, the dependence of the frequency of oscillation with $\tau_p$ is in qualitative agreement with the results in Fig.~\ref{spectrum}(c) and (d); note that as $\textrm{St}$ increases, the main peak of the spectrum moves from $\omega \approx N$ to lower frequencies (as a reference, for $\textrm{St}=3$ Eq.~(\ref{ec.osc}) yields a frequency $\Omega \approx 0.82 N$).

\begin{figure}
\includegraphics[width=1\textwidth]{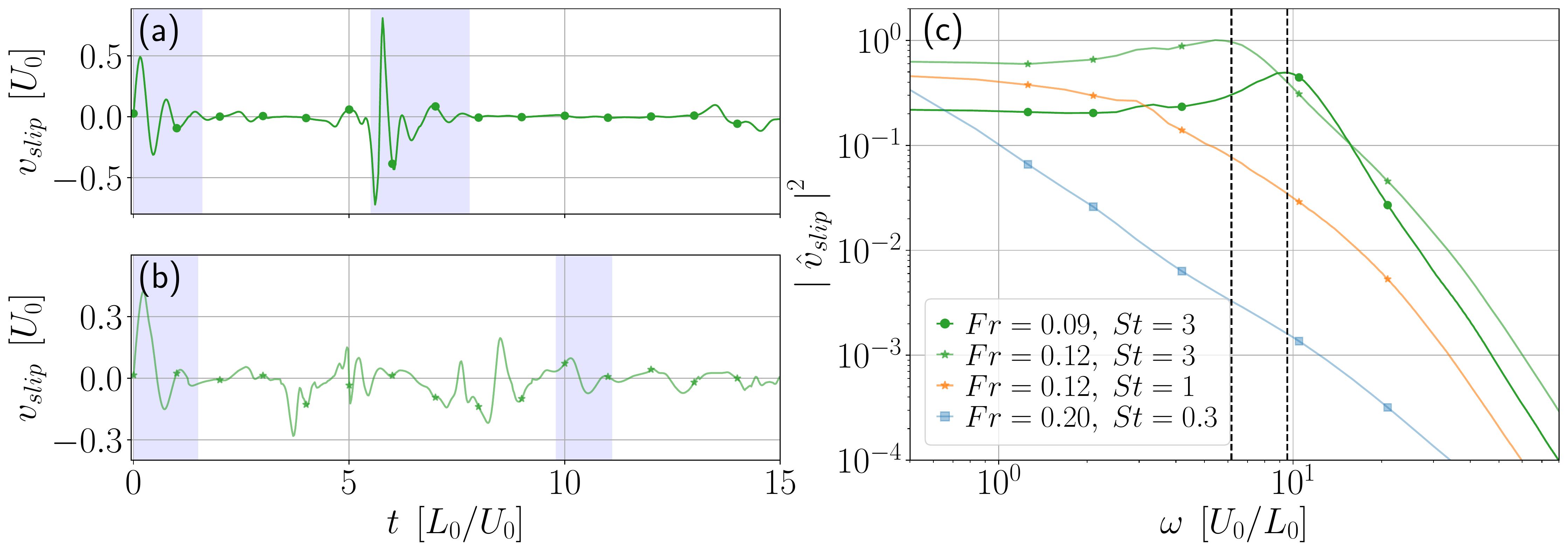}
\caption{Time series of the vertical slip velocity for two particles with $\textrm{St} = 3$ in stratified fluids with (a) $\textrm{Fr}= 0.09$ and (b) $\textrm{Fr} = 0.12$. Shaded regions indicate motions reminiscent of underdamped oscillations. (c) Power spectrum of the particles' vertical slip velocity, for different values of $\textrm{Fr}$ and $\textrm{St}$ (see inset). Overdamped particles show no peak in the spectrum, while underdamped particles display a peak at frequencies indicated by the vertical dashed lines.}
\label{slip}
\end{figure}

Equation (\ref{ec.osc}) can be integrated numerically if $f_{wav}$ is prescribed. As we do not know the precise evolution of $u_z$ as seen by each particle, we assume $u_z$ is a random colored process. The spectrum of the fluid vertical velocity in many stably stratified flows is compatible with the Garrett-Munk spectrum, as observed in oceanic observations \cite{D'Asaro_2007} and in numerical simulations \cite{Sujo_lag_2019}. This is a flat power spectrum for frequencies $\omega \le N$, resulting from the a superposition of internal gravity waves, followed by a power law decay for $\omega>N$. Thus, we consider a random superposition of oscillators of the form
\begin{equation}
z_{wav} = u_0 \, {\mathcal Re}  \left( \sum_{\omega}^{N \ge \omega>N/4}\frac{e^{i\omega t + \phi_{\omega}}}{\omega} + \sum_{\omega}^{\omega > N} N \frac{e^{i\omega t+ \phi_{\omega}}}{\omega^2}\right),
\label{ec.osc}
\end{equation}
where $\phi_{\omega}$ are random phases (note that, as we are interested only in vertical motions, the dependence of traveling waves on $x$ and $y$ can be ignored or absorbed into the random phases), and $u_0$ is an amplitude chosen so that $\dot{z}_{wav}$ has the same r.m.s.~value as that of $u_z$ in the numerical simulations. The power spectrum of $\dot{z}_{wav}$ that results is compatible with oceanic observations of the Garret-Munk spectrum \cite{dasaro_2000, D'Asaro_2007}. In other words, this process results in $\dot{z}_{wav}$ being a random variable compatible with that spectrum.

Figure \ref{model_spectrum} shows the power spectrum obtained after integrating Eq.~(\ref{ec.osc}) using this random process as the forcing, for different values of the Brunt-Väisälä frequency and the Stokes number. Spectra are qualitatively similar to those shown in Fig.~\ref{spectrum}. The spectra display two peaks, the main one close to $\omega \approx N$. At fixed $\textrm{Fr}$, increasing $\textrm{St}$ results in a broadening of this peak towards smaller frequencies. The second peak at lower frequencies has increasing amplitude with decreasing $\textrm{Fr}$, and appears at similar frequencies as those in Fig.~\ref{spectrum}. It is natural to ask whether these peaks are caused by the forcing or by the damped oscillations of the particles. Changing the forcing while still maintaining the Garret-Munk spectrum for $u_z$ (e.g., setting $f_{wav} = \ddot{z}_{wav}$) yields the same qualitative results, which indicates the peaks in the spectra are partially associated to the damped dynamics of the particles.

\begin{figure}
\includegraphics[width=0.46\textwidth]{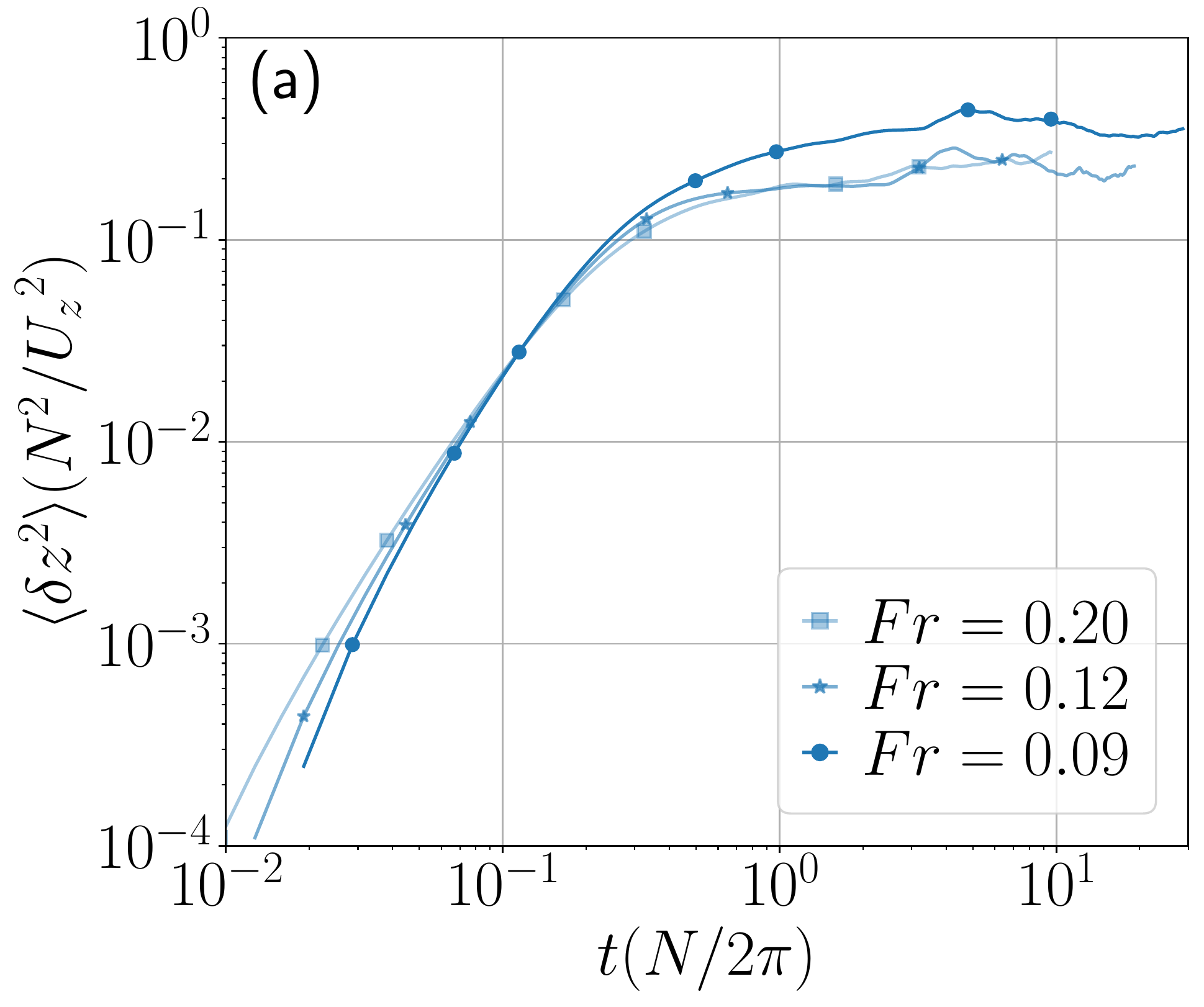}
\includegraphics[width=0.46\textwidth]{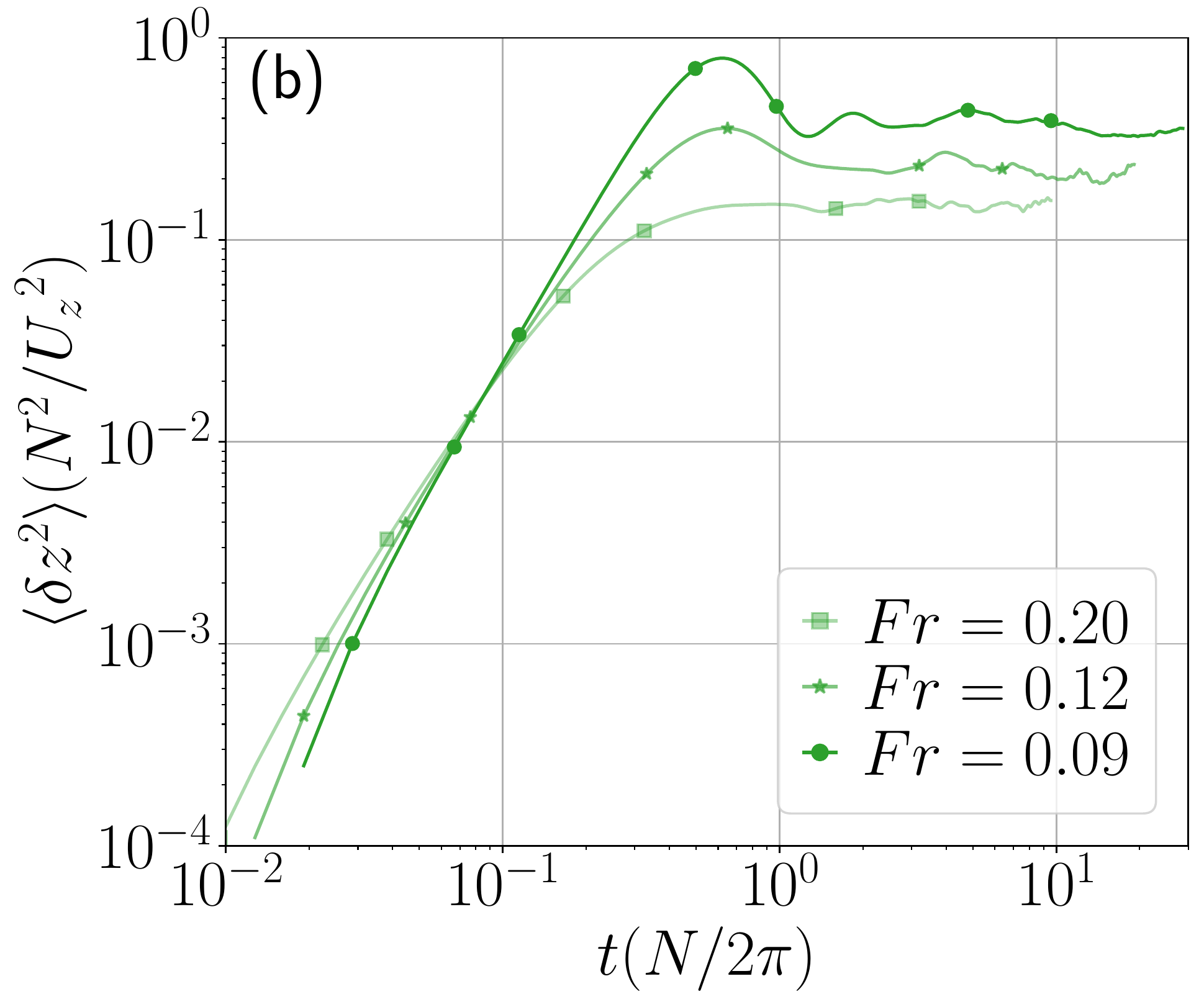}
\includegraphics[width=0.46\textwidth]{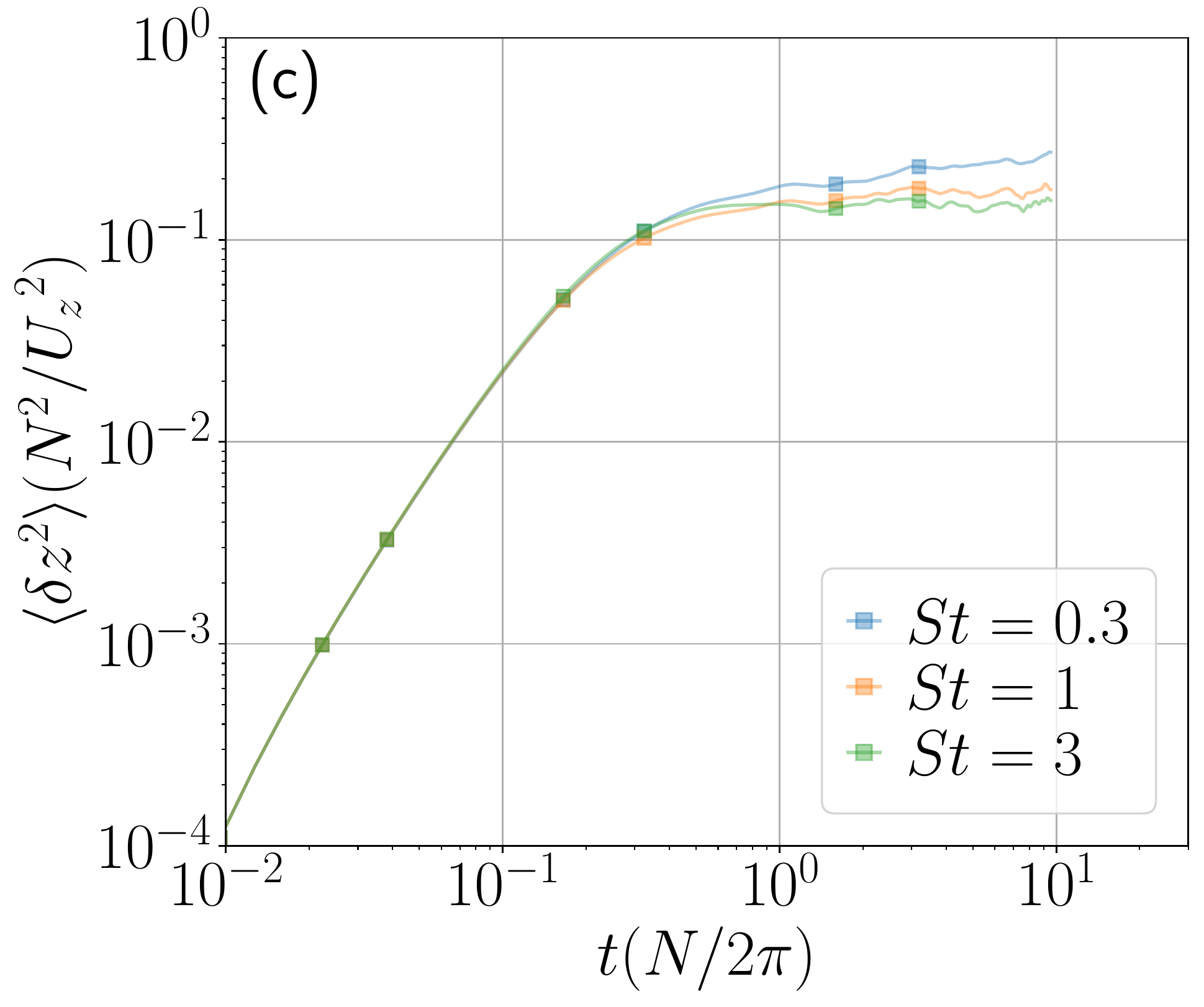}
\includegraphics[width=0.46\textwidth]{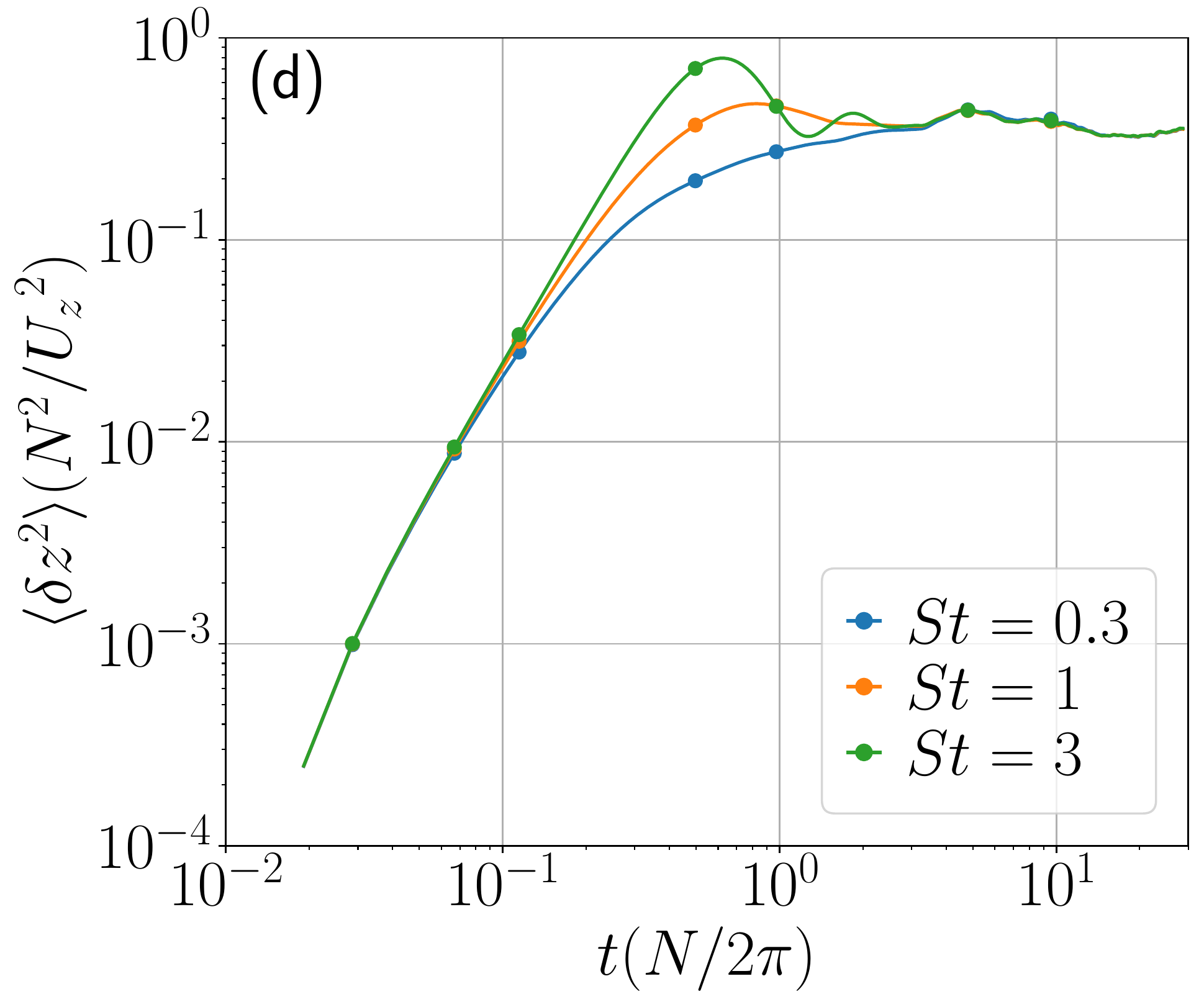}
\caption{Mean squared vertical displacements $\left< \delta z^2 \right>$ as a function of time for different values of $\textrm{Fr}$ and for (a) $\textrm{St}=0.3$ and (b) $\textrm{St} = 3$. Also, for different values of $\textrm{St}$ and for (c) $\textrm{Fr} = 0.20$ and (d) $\textrm{Fr} = 0.09$. In all cases, $\delta z^2$ in normalized by $U^2_z/N^2$, and time is normalized by $2\pi/N$.}
\label{disp}
\end{figure}

Equation (\ref{ec.osc}) can be also rewritten in terms of the vertical slip velocity, $v_{slip} = u_z-v_z$. Taking $y = z - z_{wav}$, Eq.~(\ref{ec.osc}) results in the homogeneous damped harmonic oscillator equation. As before, the resulting oscillation frequency is $\Omega^2 = 2N^2/3 - (2\tau_p)^{-2}$, with exponential decay rate $(2\tau_p)^{-1}$. Noting that $\dot{y} = v_{slip}$, we can expect the vertical slip velocity of the particles to display overdamped or underdamped oscillations depending on the sign of $\Omega^2$. Figure \ref{slip} shows $v_{slip}$ for particles in the numerical simulations with $\textrm{St} = 3$ ($\tau_p = 0.235 T_0$) in an stratified fluid with $N=12/T_0$ and $8/T_0$. Both cases have $\Omega^2 > 0$, and dynamics reminiscent of  underdamped oscillations can be identified in the time series. The power spectrum of $v_{slip}$ for multiple simulations, also shown in Fig.~\ref{slip}, shows peaks at the expected value of $\Omega$ in these two simulations, and no peaks in the other simulations with $\Omega^2<0$. Thus, the dynamics of the individual particles is compatible with randomly forced damped oscillators, with both $\tau_p$ and $N$ controlling the particles' dynamical regime.

\begin{figure}
\includegraphics[width=0.46\textwidth]{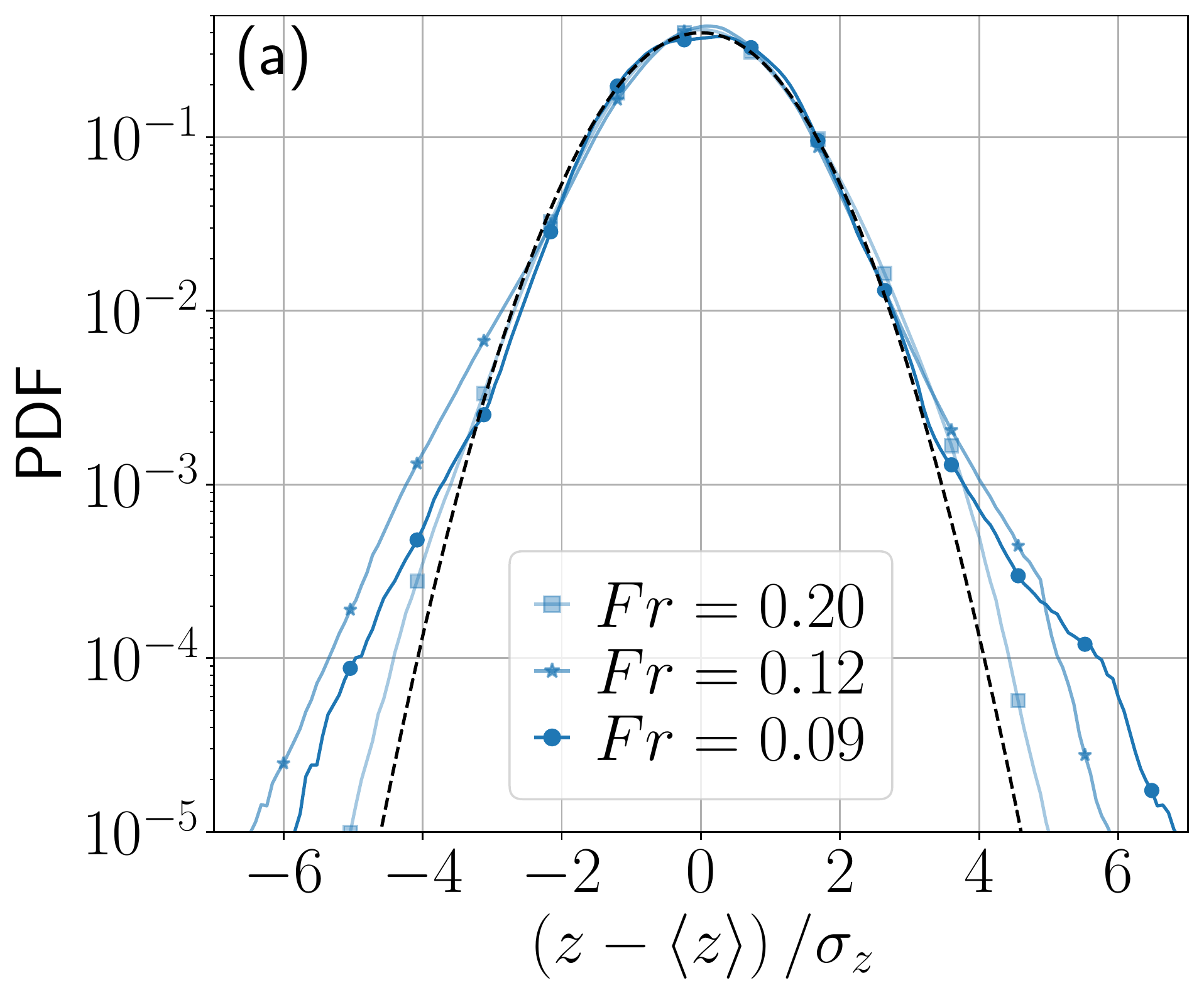}
\includegraphics[width=0.46\textwidth]{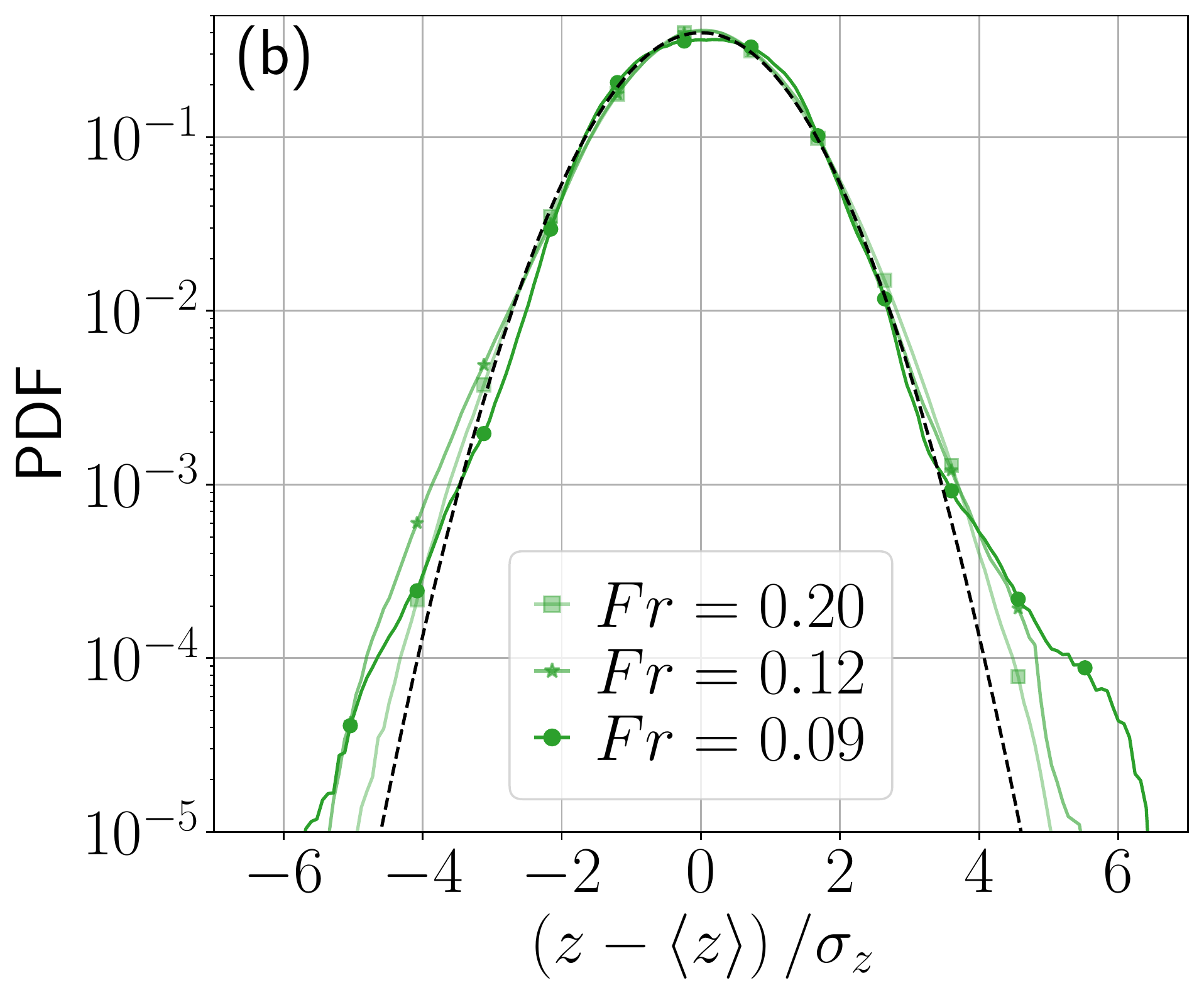}
\includegraphics[width=0.46\textwidth]{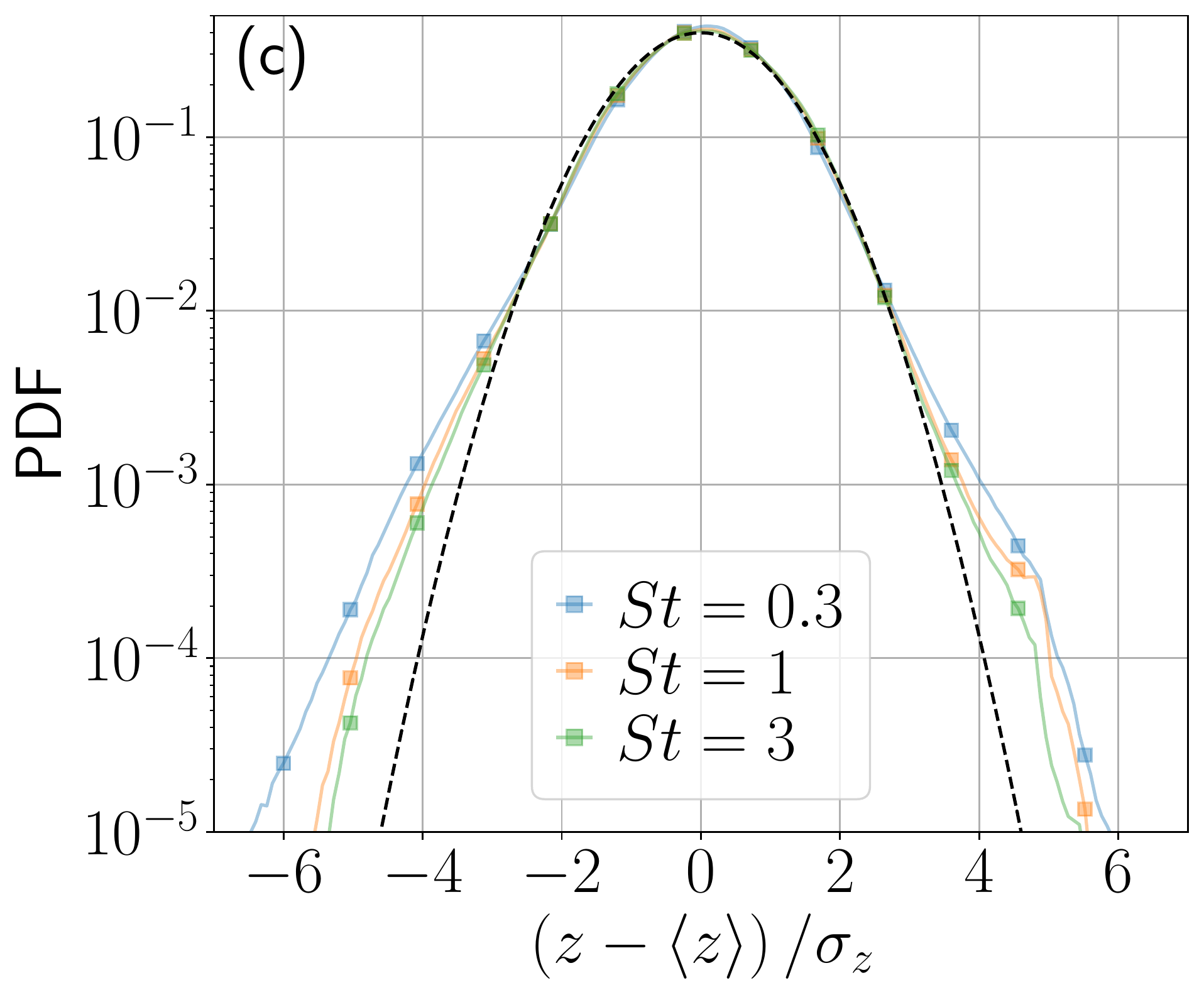}
\includegraphics[width=0.46\textwidth]{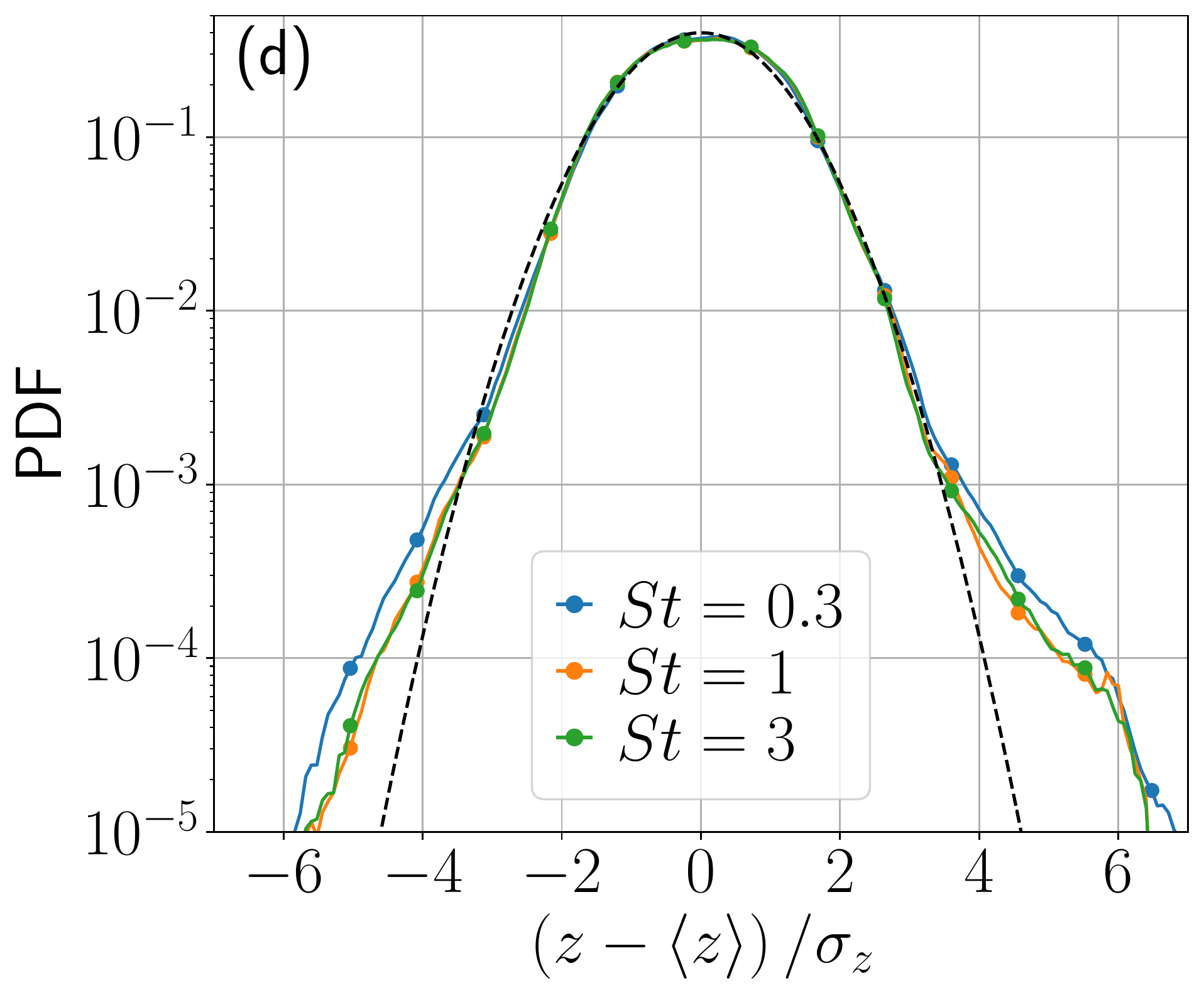}
\caption{Normalized probability density functions (PDFs) of the $z$ position of particles for different values of $\textrm{Fr}$ and (a) $\textrm{St}=0.3$ and (b) $\textrm{St} = 3$. Same for different values of $\textrm{St}$ and (c) $\textrm{Fr} = 0.20$ and (d) $\textrm{Fr} = 0.09$. The black dashed lines indicate as a reference a normal distribution.}
\label{his_z}
\end{figure}

\section{Vertical dispersion of inertial particles} \label{sec:dispersion}

Stratification limits vertical motions of the particles, strongly impairing vertical dispersion, and resulting in saturation of the mean squared vertical displacements of the particles with time. Linear models predict this saturation to take place after $t \approx 2\pi / N$, as particle displacements get constrained vertically by stratification, resulting in oscillatory motions around the neutrally buoyant equilibrium \cite{nicolleau_2000}. This was confirmed in numerical simulations with moderate $\textrm{Rb}$  \cite{lindborg_2008}. Later, studies of vertical dispersion of tracers in stably stratified flows \cite{van_aartrijk_2008, Sujo_lag_2019}, and of small neutrally-buoyant particles with small $\textrm{St}$ \cite{van_aartrijk_2010}, explicitly confirmed the saturation of the mean squared vertical dispersion. For neutrally-buoyant inertial particles the saturation was found to be faster and stronger than for Lagrangian tracers \cite{van_aartrijk_2010}.

\begin{figure}
\includegraphics[width=0.46\textwidth]{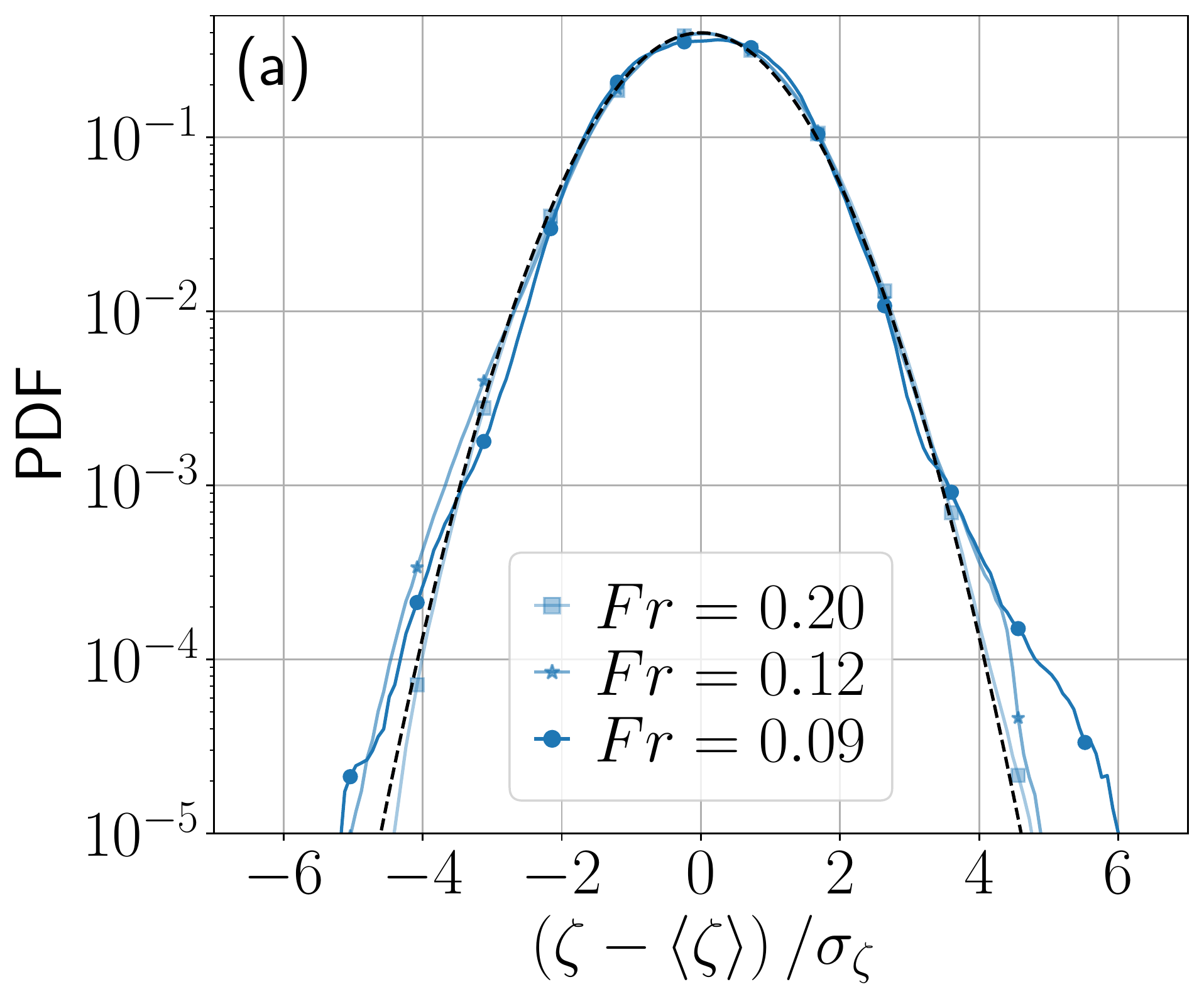}
\includegraphics[width=0.46\textwidth]{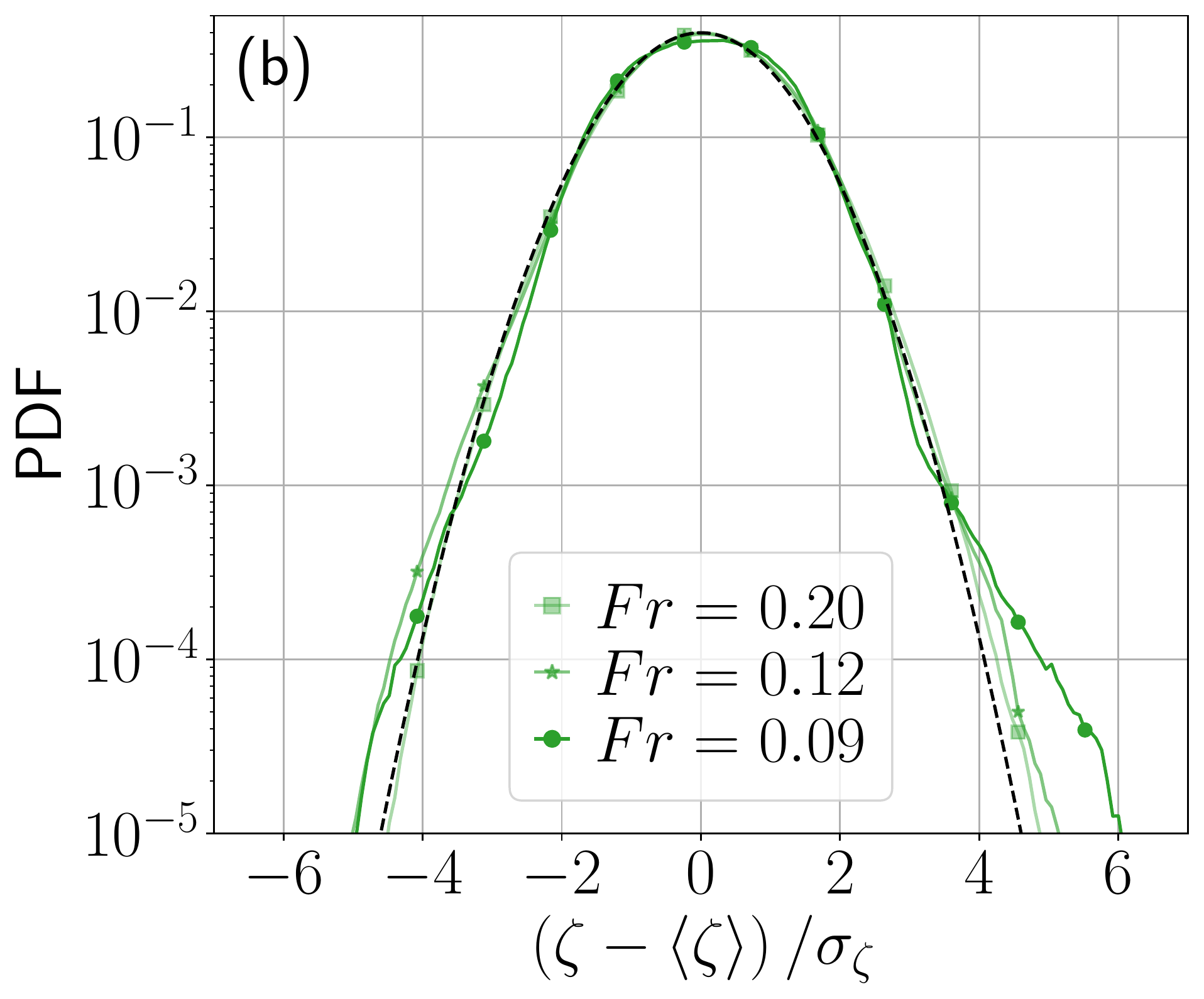}
\includegraphics[width=0.46\textwidth]{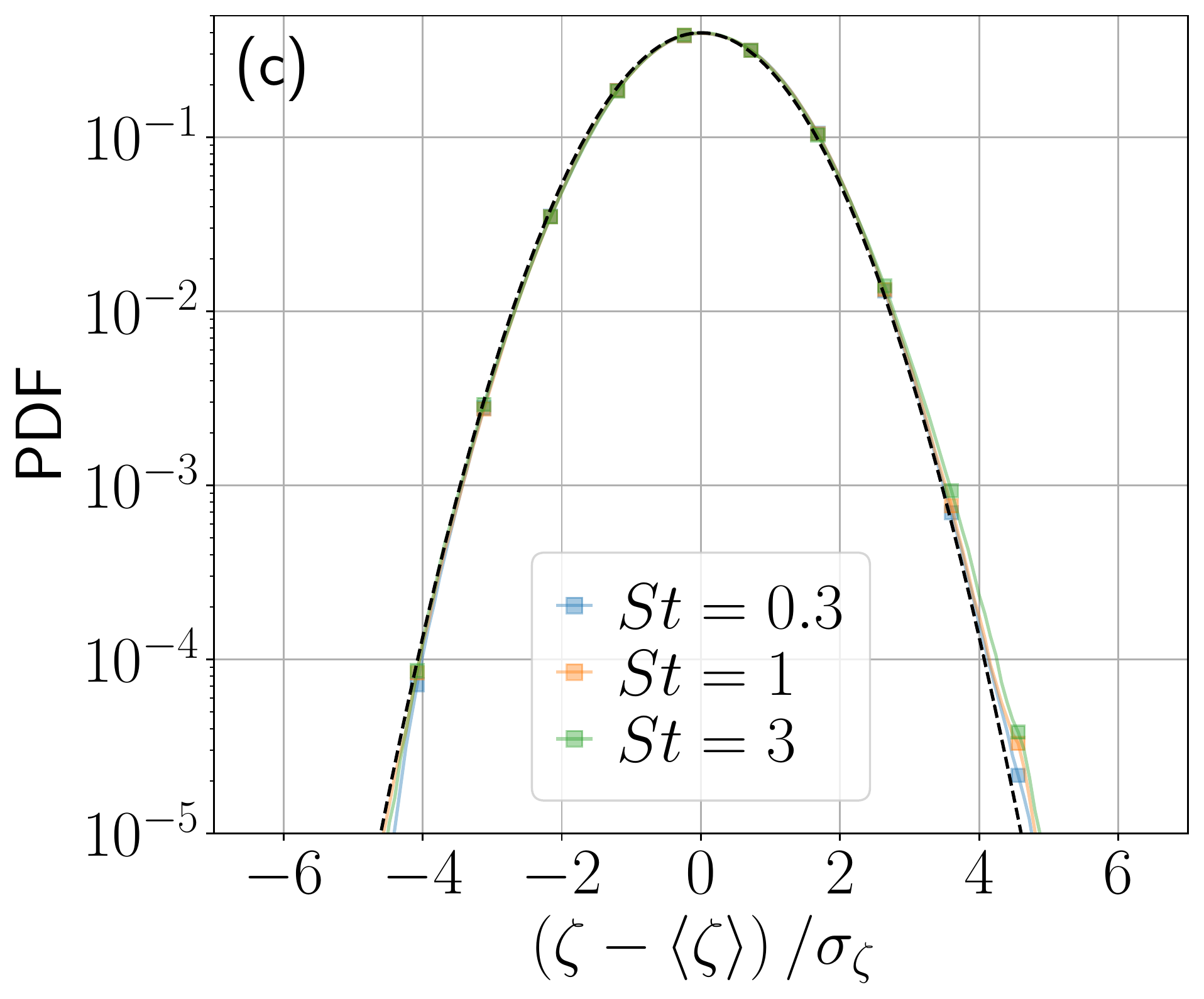}
\includegraphics[width=0.46\textwidth]{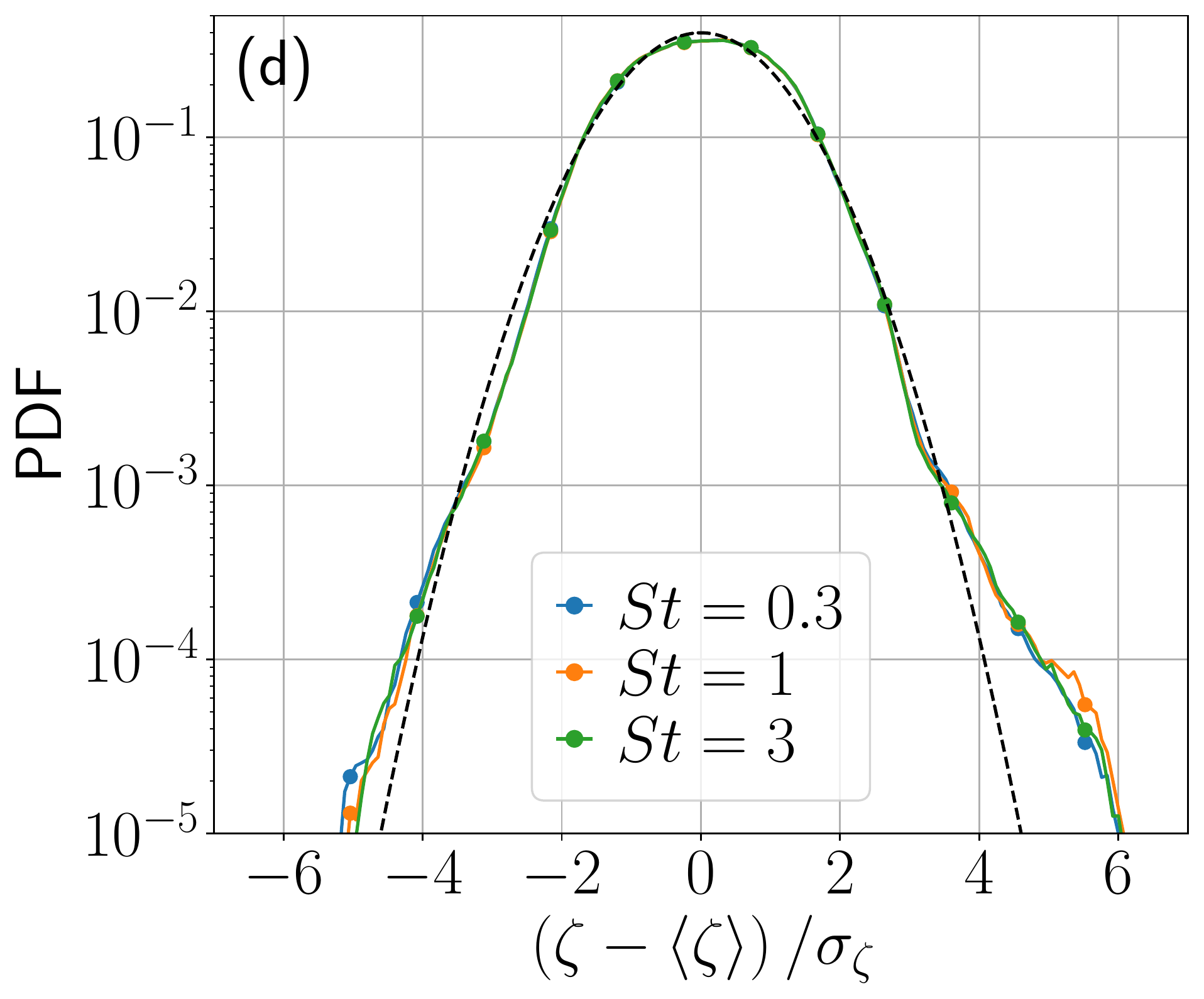}
\caption{Normalized PDFs of the rescaled density fluctuations $\zeta$ at the particles positions, in simulations with different $\textrm{Fr}$ and (a) $\textrm{St}=0.3$, (b) $\textrm{St} = 3$. Same for particles with different $\textrm{St}$ in flows with (c) $\textrm{Fr} = 0.20$ and (d) $\textrm{Fr} = 0.09$. Black dashed lines indicate a normal distribution.}
\label{histo_xi}
\end{figure}

Figure \ref{disp} shows the particles' mean squared vertical displacement in the simulations, $\left< \delta z^2(t) \right> =\left<[z_{i}(t)-z_{i}(0)]^{2}\right>_{i}$ (where the subindex $i$ indicates the average is computed over all particle labels), for different values of $\textrm{Fr}$ and $\textrm{St}$. Time is normalized by $2 \pi /N$ and $\left< \delta z^2(t) \right>$ is normalized by $(U_{z}/N)^{2} $, where $U_{z}$ is the Eulerian r.m.s.~fluid vertical velocity in the turbulent steady state. With this normalization curves collapse from $t=0$ to $t\lesssim 2\pi/N$, in a time interval with ballistic behavior. The end of this regime at a time proportional to the wave period $2\pi/N$, instead of the Lagrangian eddy turnover time, indicates that the rapid early vertical displacements are caused by the inertial particles following the inertial waves. Note also that there is more overshooting in the vertical displacements (i.e., $\langle \delta z^2 \rangle$ reaches larger values in its maximum at the end of this ballistic stage) as $\textrm{St}$ (and particle inertia) increases.  After this maximum, inertial particles oscillate around their equilibrium position, displaying a plateau in the mean squared vertical displacements, as also reported in \cite{van_aartrijk_2010}. The amplitude of the plateau is weakly dependent on $\textrm{St}$, and depends strongly on $\textrm{Fr}$ (see \cite{sm} for a movie showing the vertical displacements of the three different types of particles when $N=4/T_0$, illustrating the confinement in vertical layers). This is different from the case of tracers in stratified flows, which for sufficiently large $\textrm{Rb}$ display some slow vertical dispersion at late times caused by turbulent eddies or by diffusion \cite{Sujo_lag_2019}.

The confinement of particles around a layer can be also characterized using the probability density function (PDF) of finding a particle at a given height, either in terms of $z$, or of the density at each particle position (i.e., of how far the particle is from the equilibrium isopycnal). Figure \ref{his_z} shows the  PDF of $z$ for different $\textrm{Fr}$ and $\textrm{St}$, centered by the mean value and normalized by the dispersion. Remarkably the PDFs have asymmetric tails, the more stronger as $\textrm{Fr}$ is increased. This can be associated to the occurrence of extreme vertical drafts in stably stratified flows for values of $\textrm{Fr}$ in the range $\approx 0.05$ to $0.30$ \cite{Feraco_2018}, which can cause more frequent and larger vertical wanderings of the particles. Figure \ref{histo_xi} shows the same PDFs but in terms of the rescaled density fluctuations $\zeta$ at the particles positions, also centered by the mean value and normalized by the dispersion. These PDFs are closer to Gaussian and less sensitive to $\textrm{St}$, but still display asymmetric tails for $\textrm{Fr}=0.09$.

\begin{figure}
\includegraphics[width=0.48\textwidth]{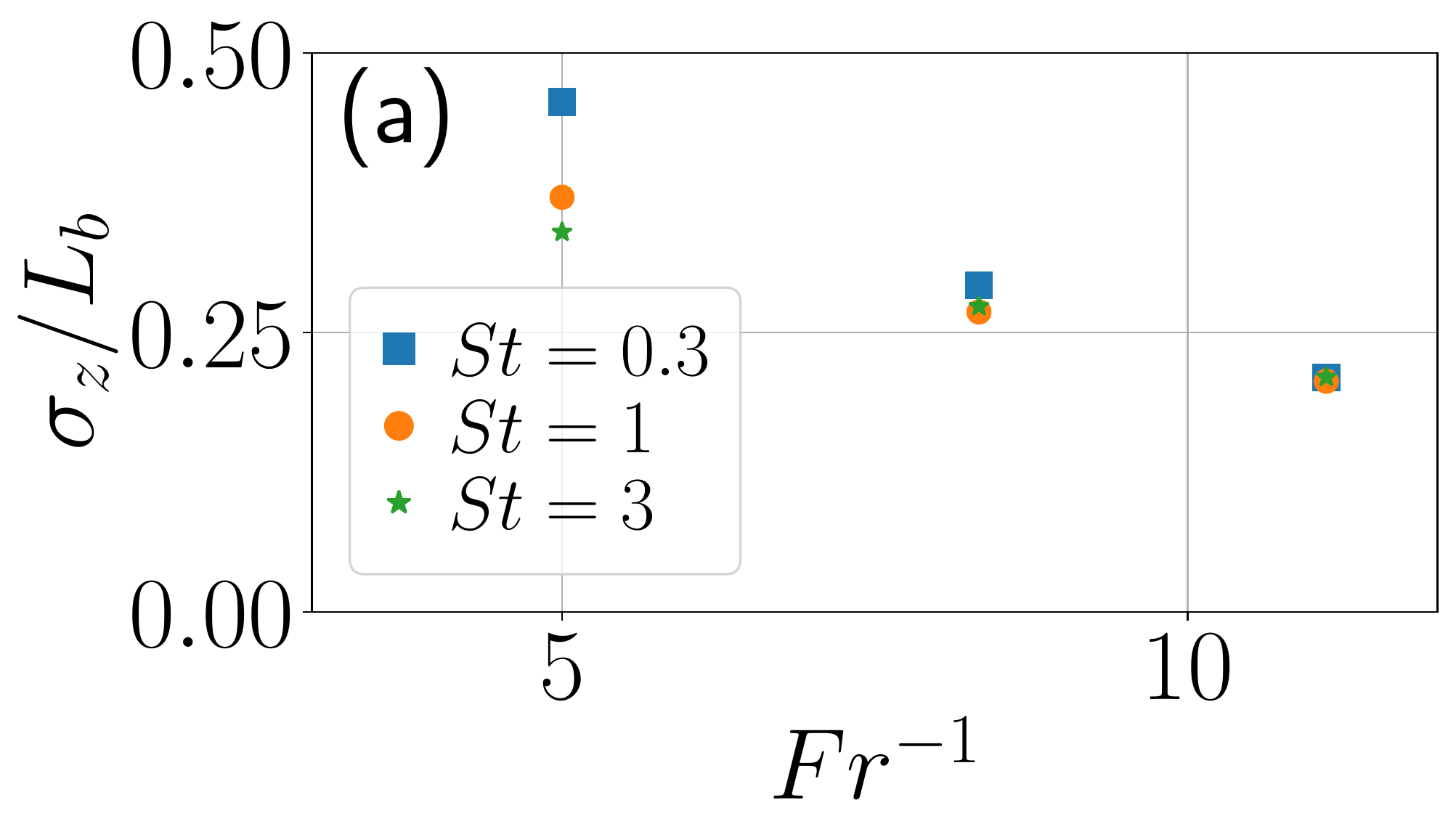}
\includegraphics[width=0.48\textwidth]{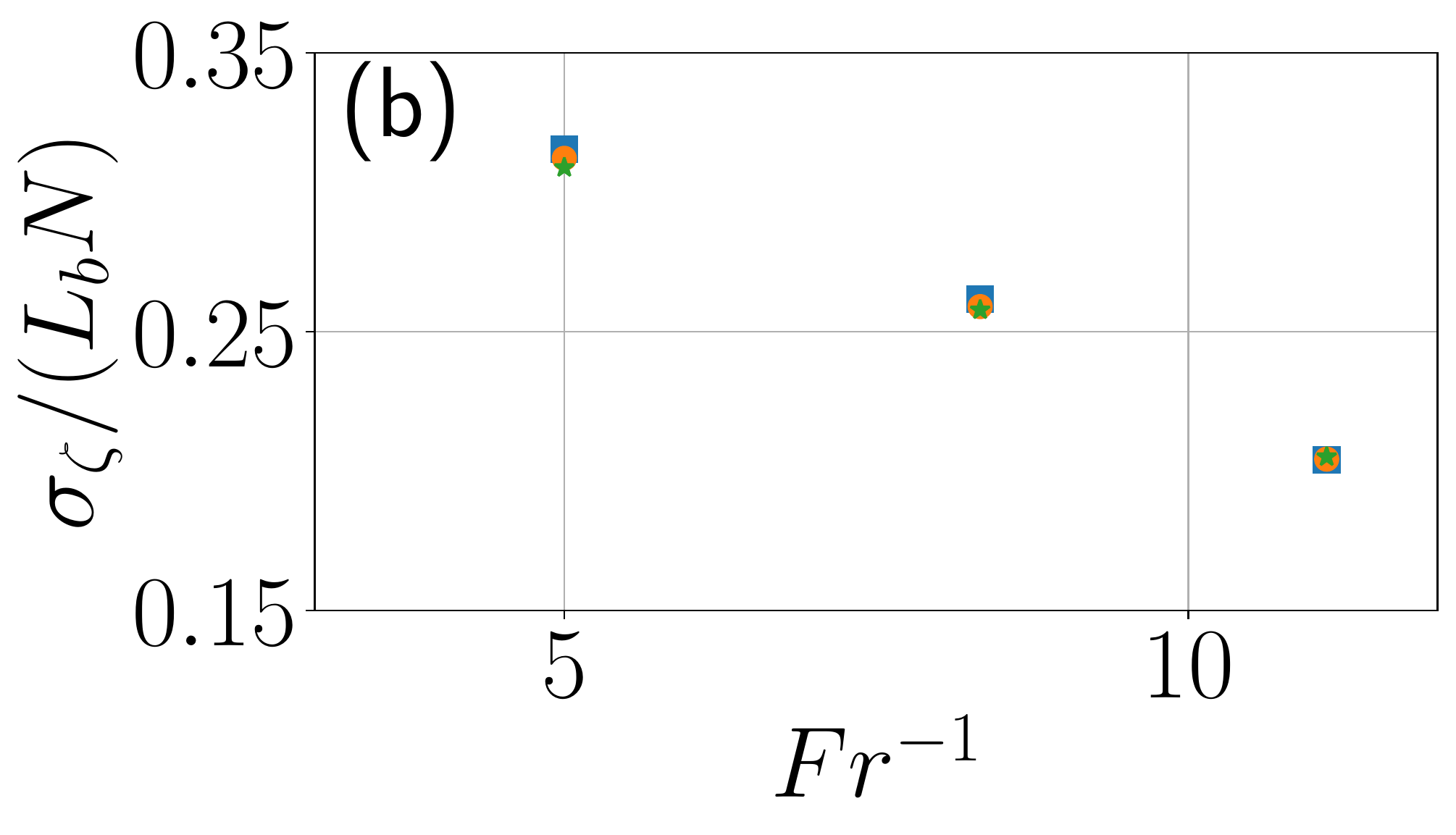}
\caption{Standard deviations of (a) the particles' positions in $z$ normalized by the buoyancy length, and (b) the fluid density variations at particles' positions, $\zeta$, normalized by the buoyancy length and the Brunt-Väisälä frequency, as a function of the inverse of the Froude number.}
\label{std}
\end{figure}

From Fig.~\ref{disp} it seems apparent that particles are confined in a narrower layer as $\textrm{Fr}$ decreases, but this is not evident from Figs.~\ref{his_z} and \ref{histo_xi} as the PDFs in those figures are normalized by their standard deviations. Figure \ref{std} show the standard deviations in $z$ and $\zeta$ of the particles, $\sigma_z$ and $\sigma_\zeta$ respectively, as a function of $\textrm{Fr}^{-1}$ for all $\textrm{St}$ considered. Note that both deviations (which can be considered as a measure of the height of the confinement layer) are a fraction of $L_b$, and decrease with decreasing $\textrm{Fr}$. The behavior of $\sigma_z$ depends also on $\textrm{St}$ for weak stratification. Sozza et al.~\cite{sozza_2018} observed that for large values of $\textrm{Fr}$, $\sigma_z$ was larger for larger $\tau_p$ (i.e., larger $\textrm{St}$), the opposite behavior of what is found here. The effect in \cite{sozza_2018} resulted from particles with more inertia being suspended from the equilibrium position for longer. Here, the mean winds in the shear layer of the Taylor-Green flow result in a different effect. Particles with more inertia are less affected by rapid vertical motions, following instead the slower horizontal motions and averaging over the vertical fluctuations as they move (see the movie in \cite{sm}). This is evident in Fig.~\ref{spectrum}, where it is observed that for the case with $\textrm{St} = 0.3$ the main peak of the power spectrum of the particles' vertical velocity is above unity, while for the case with $\textrm{St} = 3$ the peak is below unity. This confirms that particles with lower inertia are more affected by the fluid vertical displacements, while particles with larger inertia are less affected by them. Thus, the behavior of $\sigma_z$ with $\textrm{St}$ for larger $\textrm{Fr}$ is not universal, and probably dependent on the flow. However, the situation is different for $\sigma_\zeta$, as shown in Fig.~\ref{disp}(b): $\sigma_z/(L_b N) = \sigma_z/U$ seems to depend linearly on $\textrm{Fr}^{-1}$, and is independent of $\textrm{St}$, at least in the range of parameters considered. Note this amounts to the dispersion of the particles around the isopycnal decreasing linearly with increasing Brunt-V\"ais\"al\"a frequency.

\begin{figure}
\includegraphics[width=0.471\textwidth]{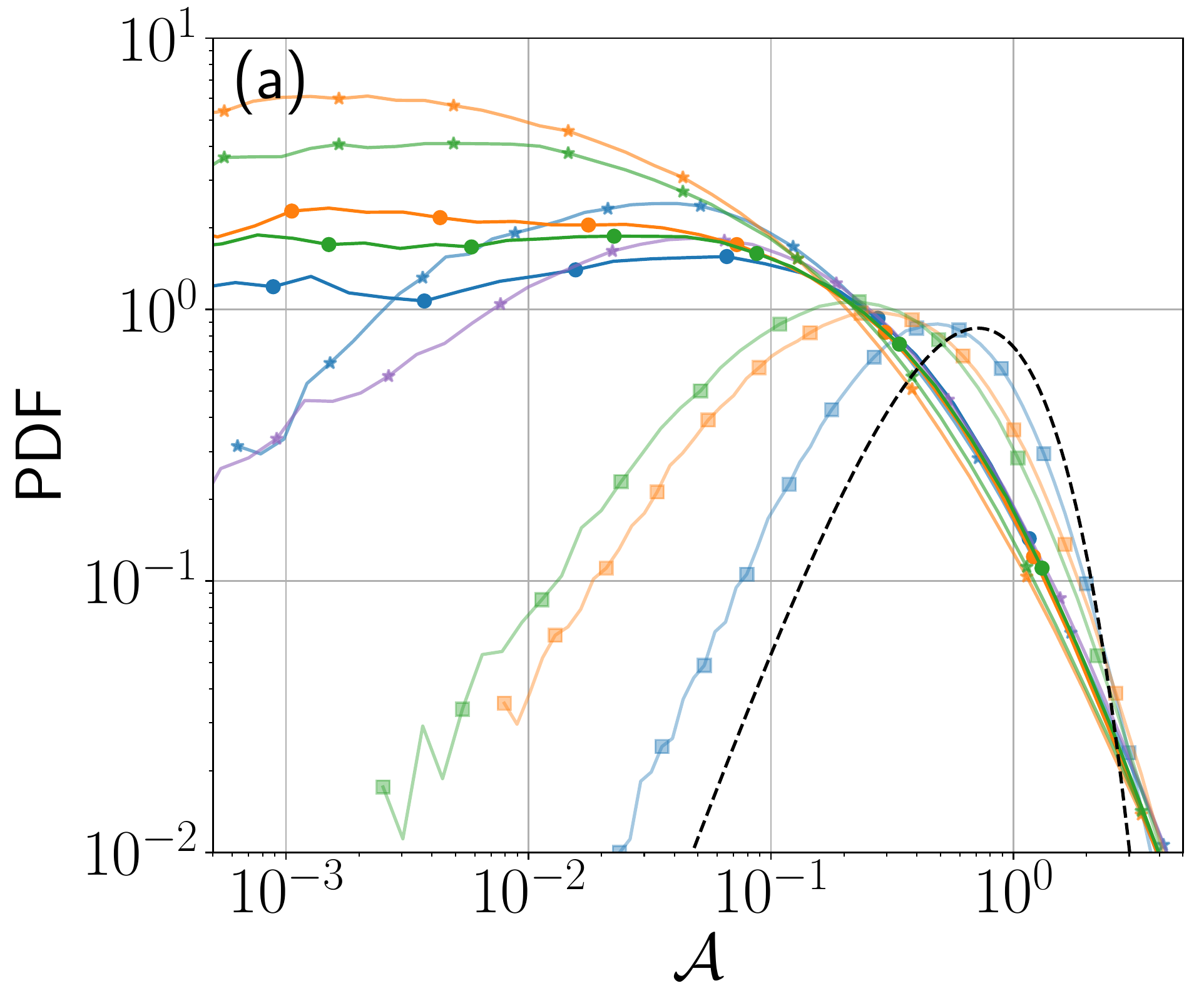}
\includegraphics[width=0.47\textwidth]{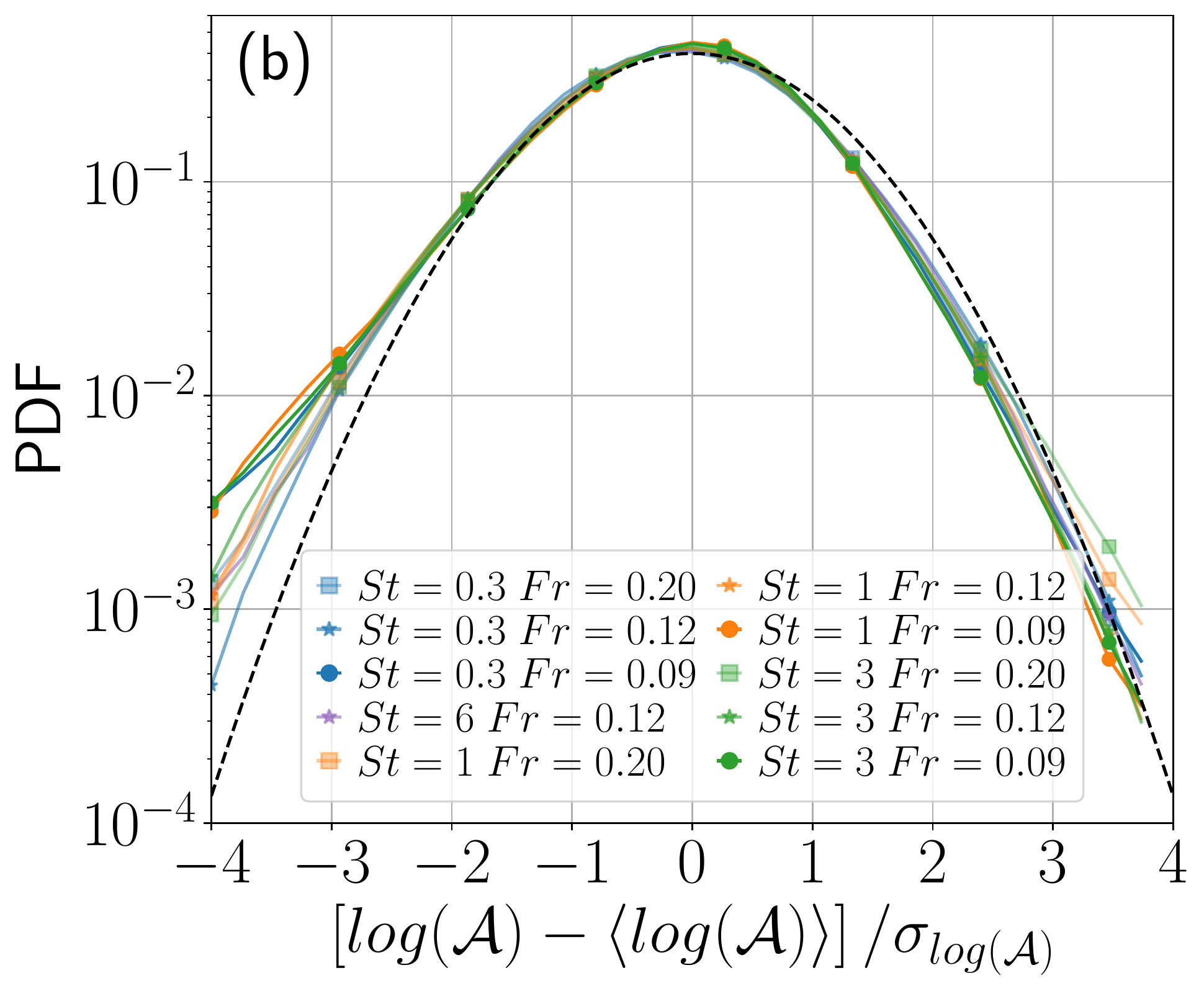}
\caption{(a) PDFs of the normalized Voronoï area $\altmathcal{A}$ for different values of $\textrm{Fr}$ and $\textrm{St}$. (b) Same PDFs centered around their mean and normalized by their dispersion. The black dashed lines indicate as a reference a random Poisson process, i.e., the PDF of randomly distributed particles.}
\label{vol}
\label{pdf_area}
\end{figure}

\section{Cluster formation and Voronoï tessellation \label{sec:cluster}}

The vertical confinement of particles have consequences for cluster formation. To quantify it we use Voronoï tessellation. Tessellations have been shown to be useful to characterize preferential concentration of particles, see, e.g., \cite{Pugliese_2022,vor15, vor16, Obligado_2015, Sumbekova_2017, Obligado_2020}, with the standard deviation of the Voronoï cell volumes or areas being associated to the amount of clustering \cite{vor15, vor16, obligado}. For heavy particles in stratified turbulence, clustering has also been studied using radial distribution functions \cite{Aartrijk_prefential_2008}, which give the ratio of the number of particle pairs found at a given separation to the expected number of pairs if particles are uniformly distributed. A Voronoï tessellation assigns a cell to each particle, so that each point in that cell is closer to that particle than to any other particle. Large tessellation cells correspond to voids (i.e., regions with far apart particles), while small cells correspond to clustered particles. While in the case of homogeneous and isotropic turbulence both three-dimensional (3D) and two-dimensional (2D) tesselations have been used, here we restrict ourselves to 2D tessellation as most of the particles remain in the thin layers discussed in the previous section. To that end, we project all particles into a plane, and consider only their $x$ and $y$ coordinates.

\begin{figure}
\includegraphics[width=0.5985\textwidth]{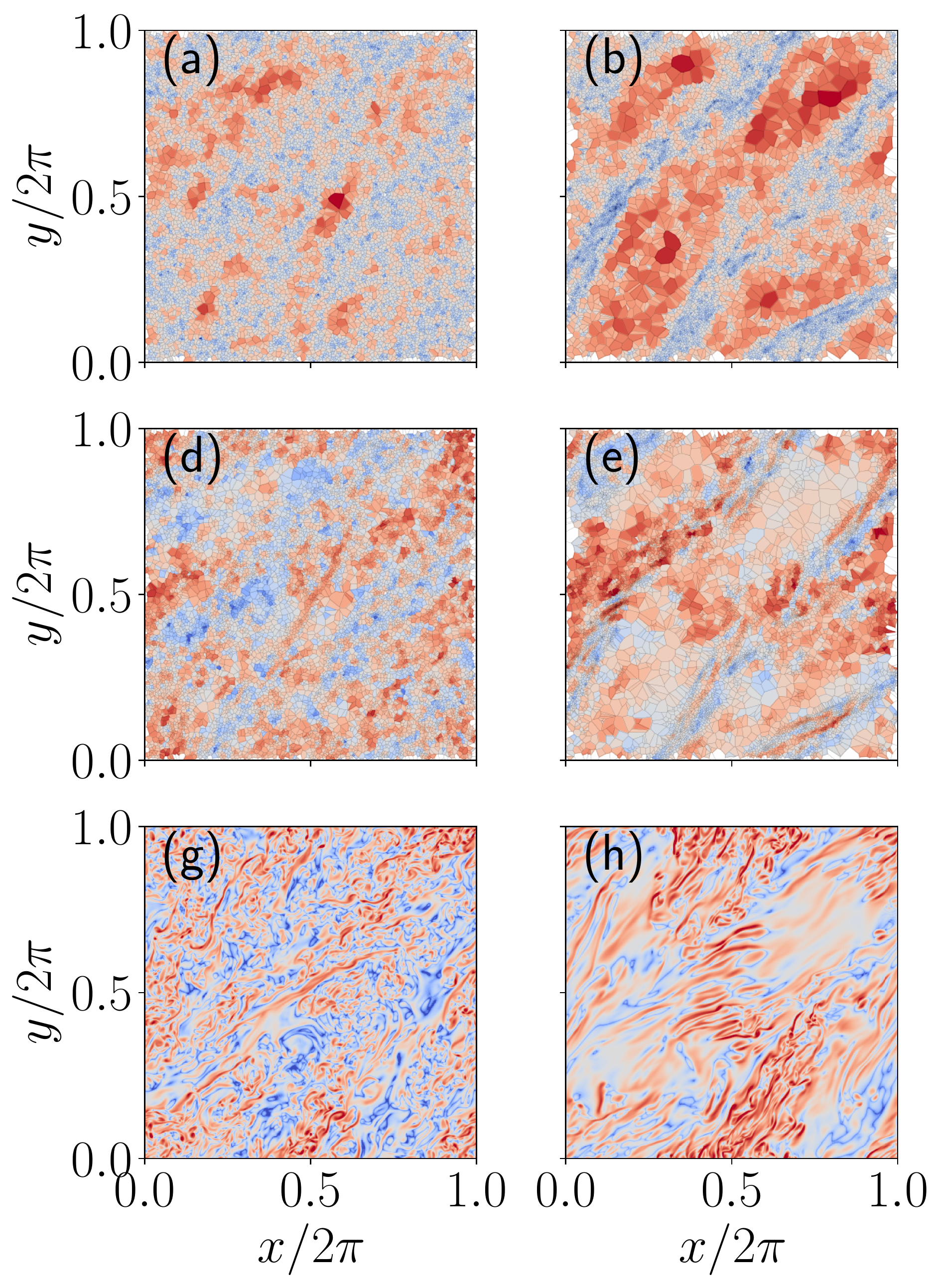}
\includegraphics[width=0.3524\textwidth]{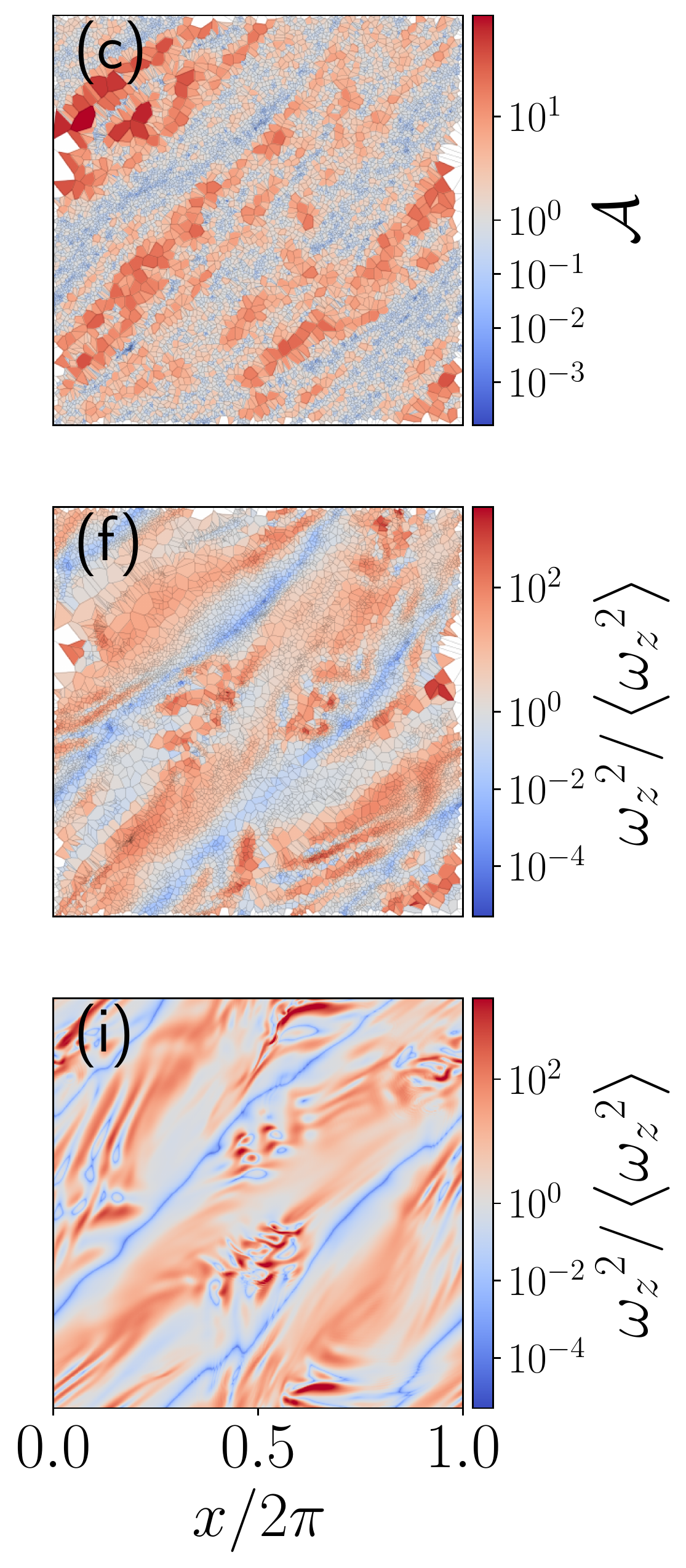}
\caption{Comparison between Voronoï particles' areas $\altmathcal{A}$ and the flow vertical vorticity. The first row shows the Voronoï areas of $10^4$ particles with $\textrm{St} = 0.3$, for (a) $\textrm{Fr} = 0.20$, (b) $0.12$, and (c) $0.09$. In the middle row, the normalized squared flow vertical vorticity $\omega_z^2$ is coarse-grained to the Voronoï areas, for the same three values of $\textrm{Fr}$, respectively in (d), (e), and (f). The bottom row shows the full resolution $\omega_z^2$ for the three values of $\textrm{Fr}$ in (g), (h), and (i).}
\label{field-voronoi}
\end{figure}

Figure \ref{pdf_area} shows the PDFs of the normalized areas of the Vorono\"i cells, $\altmathcal{A} = A/\langle A \rangle$, where $A$ is the area of each cell. The figure also shows as a reference the PDF of a random Poisson process (RPP), which corresponds to particles randomly distributed in space \cite{Tanemura_2003, Uhlmann_2020}. The first crossing from the left between the PDFs and the RPP is often used to define clusters: an excess of smaller cells are an indication of a spatial accumulation of particles in certain regions of the flow. Note that particles in the flow with $\textrm{Fr}=0.2$ are closer to the RPP. This is to be expected, as neutrally buoyant small particles do not cluster in the limit of homogeneous and isotropic turbulence (i.e., for large enough $\textrm{Fr}$) \cite{Reartes_2021}. For fixed $\textrm{Fr}$, clustering increases with $\textrm{St}$. But more importantly, clustering increases rapidly as $\textrm{Fr}$ is decreased. The strongest clustering is obtained for intermediate stratification at $\textrm{Fr} = 0.12$ and $\textrm{St} = 1$. This indicates that although the increase in stratification is favorable for cluster formation, its effect is not monotonous with $\textrm{Fr}$.

\begin{figure}
\begin{center}
\includegraphics[width=1\textwidth]{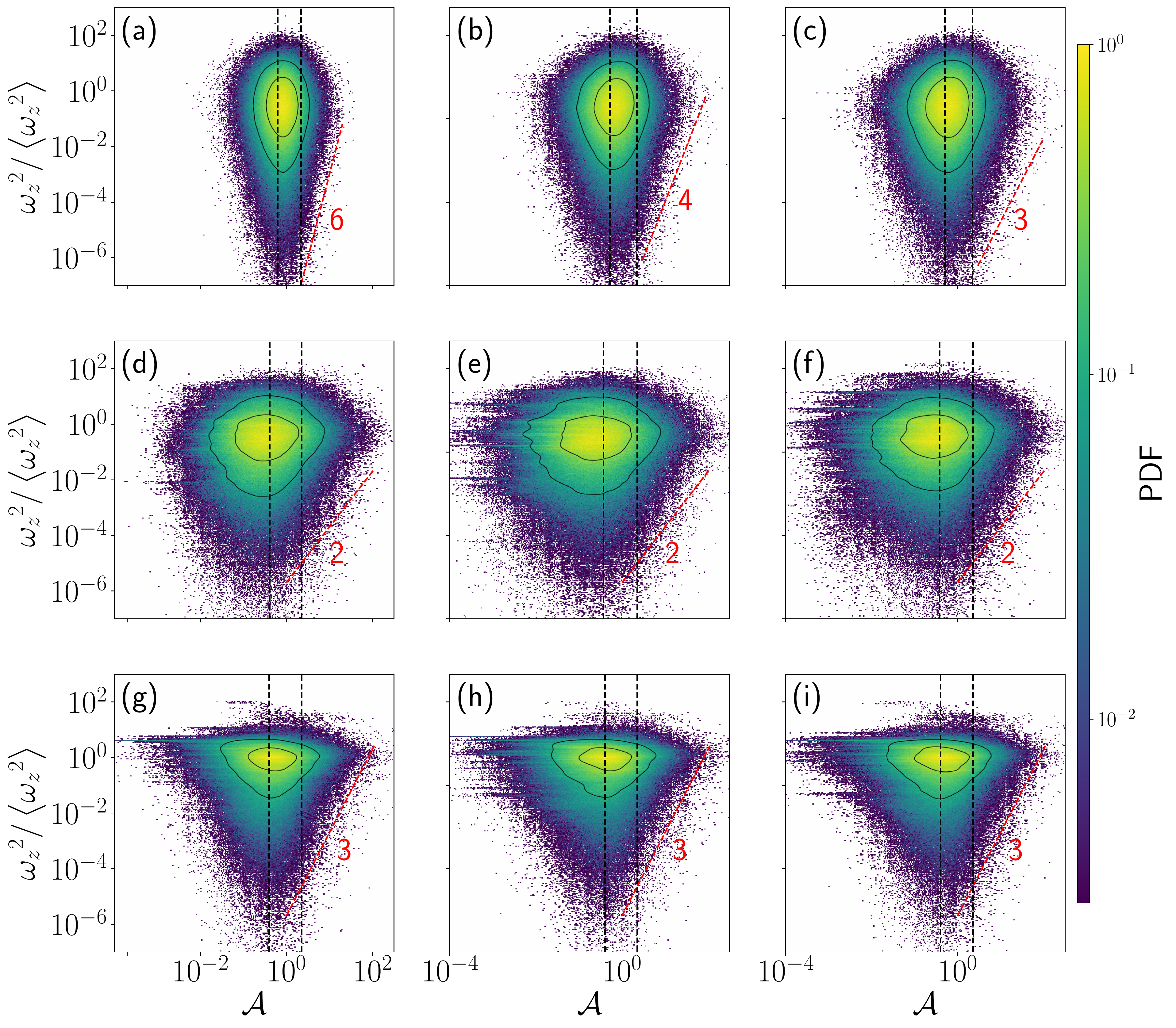}
\end{center}
\caption{Joint PDFs of $\mathcal{A}$ and ${\omega_z}^2$. From left to right, $St = 0.3$, $1$, and $3$. From top to bottom, $\textrm{Fr}=0.20$, $0.12$, and $\textrm{Fr} = 0.09$. As an example, panel (e) has $St = 0.3$ and $\textrm{Fr}=0.12$. Vertical dashed lines, from left to right, indicate respectively the first and second crossings of the PDF of $\mathcal{A}$ with the RPP. Several slopes are indicated with red dashed curves only as a reference.}
\label{a_vs_w}
\end{figure}

The clusters in this case seem to form in regions of the flow with low vertical vorticity, thus resulting from centrifugal vortex expulsion \cite{1}. To illustrate this Fig.~\ref{field-voronoi} shows the Voronoï areas of a random subset of $10^4$ particles with $\textrm{St} = 0.3$, in the three simulations with $\textrm{Fr} = 0.20$, $0.12$, and $0.09$. Red regions in panels (a), (b), and (c) correspond to cells larger than the average (voids), and blue areas to cells smaller than the average (clusters). A movie with the time evolution of the particles in the case with $\textrm{Fr} = 0.12$ can be seen in \cite{sm}. Panels (d), (e), and (f) show the squared vertical vorticity $\omega_z^2$ averaged in each Vorono\"i cell, and normalized by its mean value (as a reference, the bottom panels show the same vorticity at full resolution, i.e., not coarse-grained). Note there is some correlation between these panels: regions of low vorticity seem to correspond to smaller Voronoï areas, specially for small $\textrm{Fr}$. A similar correlation between clusters and low vorticity regions was reported before for the case of heavy particles in stratified turbulence in \cite{Aartrijk_prefential_2008}.

Figure \ref{a_vs_w} further confirms this correlation by showing joint PDFs of $\mathcal{A}$ vs ${\omega_z}^2$ for all simulations and particles. In the panels the vertical dashed lines indicate, from left to right, the first and second crossings of the PDFs of $\mathcal{A}$ with the RPP (i.e., the values of $\mathcal{A}$ below and above which cells correspond respectively to clustered particles, and to voids). Particles in voids tend to be in regions of larger vorticity, and the correlation is more clear as $\textrm{Fr}$ decreases. As a reference, we indicate different slopes with straight lines. Note that for strong stratification ($\textrm{Fr} = 0.12$), the shape of the PDFs becomes almost insensitive to the value of $\textrm{St}$. Overall, a correlation between large Vorono\"i areas and large vorticity appears independently of the Stokes number in the strongly stratified cases.

\section{Conclusions}

We presented a numerical study of the transport and spatial accumulation of light neutrally-buoyant inertial particles in stably stratified turbulent flows, using the Maxey-Riley equation for small particles. We showed that in the stratified case, the equation can be written as the equation of a driven damped oscillator, with two regimes controlled by the inverse squared particle response time, $\tau_p^{-2}$, and the flow Brunt-V\"ais\"al\"a frequency, $N$. When the former is larger particles are overdamped, while when the latter is larger particles are underdamped. This results in the appearence of two peaks in the power spectrum of the particles' vertical velocity, the main peak with frequency $\Omega \approx [2N^2/3 - (2\tau_p)^{-2}]^{-1/2}$.

As observed in previous studies of light and heavy particles in stably stratified turbulence \cite{Aartrijk_prefential_2008, van_aartrijk_2010}, the vertical dispersion of particles is strongly confined in layers. The width of this layer depends on the Stokes and Froude numbers. However, when studied in terms of density isopycnals, the width becomes independent of the particles' Stokes number (at least, in the range of parameters considered in this study), and varies with the Brunt-V\"ais\"al\"a frequency.

This vertical confinement also has a strong impact in the clustering of particles, and in the physical mechanism behind cluster formation. We showed that a two-dimensional Vorono\"i tesselation can be used to study clusters; previous studies using other methods for vertically confined particles can be seen in \cite{Aartrijk_prefential_2008, n5}. Our analysis indicates that in sufficiently stratified flows the formation of clusters is governed by centrifugal vortex expulsion, independently of whether the Stokes number is smaller or larger than unity. Moreover, clustering is strongly enhanced as stratification is increased, and this enhancement takes place also when only considering the two-dimensional positions of the particles, i.e., independently of the particles' vertical confinement. This result can be important to compute particle collisions and particle-turbulence interactions in atmospheric problems \cite{Shaw_2003}, and in oceanic flows where patches of phytoplankton and of nutrients are commonly observed \cite{Squires_1995, Martin_2003, Durham_2013}. The result is also reminiscent of observations of clustering in floaters in free surface flows, where the large scale flow circulation can play an important role in particle accumulation \cite{Del_Grosso_2019}.

\begin{acknowledgments}
The authors acknowledge financial support from UBACyT Grant No.~20020170100508BA and PICT Grant No.~2018-4298. This research was supported in part by the National Science Foundation under Grant No.~NSF PHY-1748958.
\end{acknowledgments}

\bibliography{ms}

\begin{thebibliography}{55}%
\makeatletter
\providecommand \@ifxundefined [1]{%
 \@ifx{#1\undefined}
}%
\providecommand \@ifnum [1]{%
 \ifnum #1\expandafter \@firstoftwo
 \else \expandafter \@secondoftwo
 \fi
}%
\providecommand \@ifx [1]{%
 \ifx #1\expandafter \@firstoftwo
 \else \expandafter \@secondoftwo
 \fi
}%
\providecommand \natexlab [1]{#1}%
\providecommand \enquote  [1]{``#1''}%
\providecommand \bibnamefont  [1]{#1}%
\providecommand \bibfnamefont [1]{#1}%
\providecommand \citenamefont [1]{#1}%
\providecommand \href@noop [0]{\@secondoftwo}%
\providecommand \href [0]{\begingroup \@sanitize@url \@href}%
\providecommand \@href[1]{\@@startlink{#1}\@@href}%
\providecommand \@@href[1]{\endgroup#1\@@endlink}%
\providecommand \@sanitize@url [0]{\catcode `\\12\catcode `\$12\catcode
  `\&12\catcode `\#12\catcode `\^12\catcode `\_12\catcode `\%12\relax}%
\providecommand \@@startlink[1]{}%
\providecommand \@@endlink[0]{}%
\providecommand \url  [0]{\begingroup\@sanitize@url \@url }%
\providecommand \@url [1]{\endgroup\@href {#1}{\urlprefix }}%
\providecommand \urlprefix  [0]{URL }%
\providecommand \Eprint [0]{\href }%
\providecommand \doibase [0]{https://doi.org/}%
\providecommand \selectlanguage [0]{\@gobble}%
\providecommand \bibinfo  [0]{\@secondoftwo}%
\providecommand \bibfield  [0]{\@secondoftwo}%
\providecommand \translation [1]{[#1]}%
\providecommand \BibitemOpen [0]{}%
\providecommand \bibitemStop [0]{}%
\providecommand \bibitemNoStop [0]{.\EOS\space}%
\providecommand \EOS [0]{\spacefactor3000\relax}%
\providecommand \BibitemShut  [1]{\csname bibitem#1\endcsname}%
\let\auto@bib@innerbib\@empty
\bibitem [{\citenamefont {Wyngaard}(1992)}]{wyngaard_1992}%
  \BibitemOpen
  \bibfield  {author} {\bibinfo {author} {\bibfnamefont {J.}~\bibnamefont
  {Wyngaard}},\ }\bibfield  {title} {\bibinfo {title} {Turbulence in the
  atmosphere},\ }\href@noop {} {\bibfield  {journal} {\bibinfo  {journal}
  {Physical Review Fluids}\ }\textbf {\bibinfo {volume} {24}},\ \bibinfo
  {pages} {205} (\bibinfo {year} {1992})}\BibitemShut {NoStop}%
\bibitem [{\citenamefont {D'Asaro}\ and\ \citenamefont
  {Lien}(2000)}]{dasaro_2000}%
  \BibitemOpen
  \bibfield  {author} {\bibinfo {author} {\bibfnamefont {E.}~\bibnamefont
  {D'Asaro}}\ and\ \bibinfo {author} {\bibfnamefont {R.-C.}\ \bibnamefont
  {Lien}},\ }\bibfield  {title} {\bibinfo {title} {Lagrangian measurements of
  waves and turbulence in stratified flows},\ }\href@noop {} {\bibfield
  {journal} {\bibinfo  {journal} {J. Phys. Oceanogr.}\ }\textbf {\bibinfo
  {volume} {20}},\ \bibinfo {pages} {641} (\bibinfo {year} {2000})}\BibitemShut
  {NoStop}%
\bibitem [{\citenamefont {Watanabe}\ \emph {et~al.}(2016)\citenamefont
  {Watanabe}, \citenamefont {Riley}, \citenamefont {de~Bruyn~Kops},
  \citenamefont {Diamessis},\ and\ \citenamefont {Zhou}}]{watanabe_2017}%
  \BibitemOpen
  \bibfield  {author} {\bibinfo {author} {\bibfnamefont {T.}~\bibnamefont
  {Watanabe}}, \bibinfo {author} {\bibfnamefont {J.}~\bibnamefont {Riley}},
  \bibinfo {author} {\bibfnamefont {S.}~\bibnamefont {de~Bruyn~Kops}}, \bibinfo
  {author} {\bibfnamefont {P.}~\bibnamefont {Diamessis}},\ and\ \bibinfo
  {author} {\bibfnamefont {Q.}~\bibnamefont {Zhou}},\ }\bibfield  {title}
  {\bibinfo {title} {Turbulent/non-turbulent interfaces in wakes in stably
  stratified fluids},\ }\href@noop {} {\bibfield  {journal} {\bibinfo
  {journal} {Journal of Fluid Mechanics}\ }\textbf {\bibinfo {volume} {797}},\
  \bibinfo {pages} {R1} (\bibinfo {year} {2016})}\BibitemShut {NoStop}%
\bibitem [{\citenamefont {Amir}\ \emph {et~al.}(2016)\citenamefont {Amir},
  \citenamefont {Bar}, \citenamefont {Eidelman}, \citenamefont {Elperin},
  \citenamefont {Kleeorin},\ and\ \citenamefont {Rogachevskii}}]{amir_2017}%
  \BibitemOpen
  \bibfield  {author} {\bibinfo {author} {\bibfnamefont {G.}~\bibnamefont
  {Amir}}, \bibinfo {author} {\bibfnamefont {N.}~\bibnamefont {Bar}}, \bibinfo
  {author} {\bibfnamefont {A.}~\bibnamefont {Eidelman}}, \bibinfo {author}
  {\bibfnamefont {T.}~\bibnamefont {Elperin}}, \bibinfo {author} {\bibfnamefont
  {N.}~\bibnamefont {Kleeorin}},\ and\ \bibinfo {author} {\bibfnamefont
  {I.}~\bibnamefont {Rogachevskii}},\ }\bibfield  {title} {\bibinfo {title}
  {Turbulent thermal diffusion in strongly stratified turbulence: Theory and
  experiments},\ }\href@noop {} {\bibfield  {journal} {\bibinfo  {journal}
  {Physical Review Fluids}\ }\textbf {\bibinfo {volume} {2}},\ \bibinfo {pages}
  {064605} (\bibinfo {year} {2016})}\BibitemShut {NoStop}%
\bibitem [{\citenamefont {Maxey}\ and\ \citenamefont {Riley}(1983)}]{1}%
  \BibitemOpen
  \bibfield  {author} {\bibinfo {author} {\bibfnamefont {M.}~\bibnamefont
  {Maxey}}\ and\ \bibinfo {author} {\bibfnamefont {J.}~\bibnamefont {Riley}},\
  }\bibfield  {title} {\bibinfo {title} {Equation of motion for a small rigid
  sphere in a nonuniform flow},\ }\href@noop {} {\bibfield  {journal} {\bibinfo
   {journal} {Physics of {F}luids}\ }\textbf {\bibinfo {volume} {26}},\
  \bibinfo {pages} {883} (\bibinfo {year} {1983})}\BibitemShut {NoStop}%
\bibitem [{\citenamefont {Obligado}\ \emph {et~al.}(2015)\citenamefont
  {Obligado}, \citenamefont {Cartellier},\ and\ \citenamefont
  {Bourgoin}}]{Obligado_2015}%
  \BibitemOpen
  \bibfield  {author} {\bibinfo {author} {\bibfnamefont {M.}~\bibnamefont
  {Obligado}}, \bibinfo {author} {\bibfnamefont {A.}~\bibnamefont
  {Cartellier}},\ and\ \bibinfo {author} {\bibfnamefont {M.}~\bibnamefont
  {Bourgoin}},\ }\bibfield  {title} {\bibinfo {title} {Experimental detection
  of superclusters of water droplets in homogeneous isotropic turbulence},\
  }\href@noop {} {\bibfield  {journal} {\bibinfo  {journal} {EPL (Europhysics
  Letters)}\ }\textbf {\bibinfo {volume} {112}},\ \bibinfo {pages} {54004}
  (\bibinfo {year} {2015})}\BibitemShut {NoStop}%
\bibitem [{\citenamefont {Tavanashad}\ \emph {et~al.}(2021)\citenamefont
  {Tavanashad}, \citenamefont {Passalacqua},\ and\ \citenamefont
  {Subramaniam}}]{Tavanashad_2021}%
  \BibitemOpen
  \bibfield  {author} {\bibinfo {author} {\bibfnamefont {V.}~\bibnamefont
  {Tavanashad}}, \bibinfo {author} {\bibfnamefont {A.}~\bibnamefont
  {Passalacqua}},\ and\ \bibinfo {author} {\bibfnamefont {S.}~\bibnamefont
  {Subramaniam}},\ }\bibfield  {title} {\bibinfo {title} {Particle-resolved
  simulation of freely evolving particle suspensions: Flow physics and
  modeling},\ }\href@noop {} {\bibfield  {journal} {\bibinfo  {journal}
  {International Journal of Multiphase Flow}\ }\textbf {\bibinfo {volume}
  {135}},\ \bibinfo {pages} {103533} (\bibinfo {year} {2021})}\BibitemShut
  {NoStop}%
\bibitem [{\citenamefont {Wagner}\ \emph {et~al.}(2019)\citenamefont {Wagner},
  \citenamefont {R{\"u}hs}, \citenamefont {Schwarzkopf}, \citenamefont
  {Koszalka},\ and\ \citenamefont {Biastoch}}]{Wagner_2019}%
  \BibitemOpen
  \bibfield  {author} {\bibinfo {author} {\bibfnamefont {P.}~\bibnamefont
  {Wagner}}, \bibinfo {author} {\bibfnamefont {S.}~\bibnamefont {R{\"u}hs}},
  \bibinfo {author} {\bibfnamefont {F.~U.}\ \bibnamefont {Schwarzkopf}},
  \bibinfo {author} {\bibfnamefont {I.~M.}\ \bibnamefont {Koszalka}},\ and\
  \bibinfo {author} {\bibfnamefont {A.}~\bibnamefont {Biastoch}},\ }\bibfield
  {title} {\bibinfo {title} {Can lagrangian tracking simulate tracer spreading
  in a high-resolution ocean general circulation model?},\ }\href@noop {}
  {\bibfield  {journal} {\bibinfo  {journal} {Journal of Physical
  Oceanography}\ }\textbf {\bibinfo {volume} {49}},\ \bibinfo {pages} {1141}
  (\bibinfo {year} {2019})}\BibitemShut {NoStop}%
\bibitem [{\citenamefont {Palmer}(2019)}]{Palmer_2019}%
  \BibitemOpen
  \bibfield  {author} {\bibinfo {author} {\bibfnamefont {T.}~\bibnamefont
  {Palmer}},\ }\bibfield  {title} {\bibinfo {title} {Stochastic weather and
  climate models},\ }\href@noop {} {\bibfield  {journal} {\bibinfo  {journal}
  {Nature Reviews Physics}\ }\textbf {\bibinfo {volume} {1}},\ \bibinfo {pages}
  {463} (\bibinfo {year} {2019})}\BibitemShut {NoStop}%
\bibitem [{\citenamefont {Beron-Vera}\ \emph {et~al.}(2019)\citenamefont
  {Beron-Vera}, \citenamefont {Olascoaga},\ and\ \citenamefont
  {Miron}}]{Beron_2019}%
  \BibitemOpen
  \bibfield  {author} {\bibinfo {author} {\bibfnamefont {F.~J.}\ \bibnamefont
  {Beron-Vera}}, \bibinfo {author} {\bibfnamefont {M.~J.}\ \bibnamefont
  {Olascoaga}},\ and\ \bibinfo {author} {\bibfnamefont {P.}~\bibnamefont
  {Miron}},\ }\bibfield  {title} {\bibinfo {title} {Building a maxey--riley
  framework for surface ocean inertial particle dynamics},\ }\href@noop {}
  {\bibfield  {journal} {\bibinfo  {journal} {Physics of Fluids}\ }\textbf
  {\bibinfo {volume} {31}},\ \bibinfo {pages} {096602} (\bibinfo {year}
  {2019})}\BibitemShut {NoStop}%
\bibitem [{\citenamefont {E.~Lindborg}(2008)}]{lindborg_2008}%
  \BibitemOpen
  \bibfield  {author} {\bibinfo {author} {\bibfnamefont {G.~B.}\ \bibnamefont
  {E.~Lindborg}},\ }\bibfield  {title} {\bibinfo {title} {Vertical dispersion
  by stratified turbulence},\ }\href@noop {} {\bibfield  {journal} {\bibinfo
  {journal} {J. Fluid Mech.}\ }\textbf {\bibinfo {volume} {614}},\ \bibinfo
  {pages} {303–314} (\bibinfo {year} {2008})}\BibitemShut {NoStop}%
\bibitem [{\citenamefont {Marino}\ \emph {et~al.}(2014)\citenamefont {Marino},
  \citenamefont {Mininni}, \citenamefont {Rosenberg},\ and\ \citenamefont
  {Pouquet}}]{marino_2014}%
  \BibitemOpen
  \bibfield  {author} {\bibinfo {author} {\bibfnamefont {R.}~\bibnamefont
  {Marino}}, \bibinfo {author} {\bibfnamefont {P.}~\bibnamefont {Mininni}},
  \bibinfo {author} {\bibfnamefont {D.}~\bibnamefont {Rosenberg}},\ and\
  \bibinfo {author} {\bibfnamefont {A.}~\bibnamefont {Pouquet}},\ }\bibfield
  {title} {\bibinfo {title} {Large-scale anisotropy in stably stratified
  rotating flows},\ }\href@noop {} {\bibfield  {journal} {\bibinfo  {journal}
  {Phys. Rev. E}\ }\textbf {\bibinfo {volume} {90}},\ \bibinfo {pages} {023018}
  (\bibinfo {year} {2014})}\BibitemShut {NoStop}%
\bibitem [{\citenamefont {Portwood}\ \emph {et~al.}(2019)\citenamefont
  {Portwood}, \citenamefont {de~Bruyn~Kops},\ and\ \citenamefont
  {Caulfield}}]{Portwood_2019}%
  \BibitemOpen
  \bibfield  {author} {\bibinfo {author} {\bibfnamefont {G.~D.}\ \bibnamefont
  {Portwood}}, \bibinfo {author} {\bibfnamefont {S.}~\bibnamefont
  {de~Bruyn~Kops}},\ and\ \bibinfo {author} {\bibfnamefont {C.}~\bibnamefont
  {Caulfield}},\ }\bibfield  {title} {\bibinfo {title} {Asymptotic dynamics of
  high dynamic range stratified turbulence},\ }\href@noop {} {\bibfield
  {journal} {\bibinfo  {journal} {Physical Review Letters}\ }\textbf {\bibinfo
  {volume} {122}},\ \bibinfo {pages} {194504} (\bibinfo {year}
  {2019})}\BibitemShut {NoStop}%
\bibitem [{\citenamefont {Smith}\ and\ \citenamefont
  {Waleffe}(2002)}]{smith_2002}%
  \BibitemOpen
  \bibfield  {author} {\bibinfo {author} {\bibfnamefont {L.}~\bibnamefont
  {Smith}}\ and\ \bibinfo {author} {\bibfnamefont {F.}~\bibnamefont
  {Waleffe}},\ }\bibfield  {title} {\bibinfo {title} {Generation of slow large
  scales in forced rotating stratified turbulence},\ }\href@noop {} {\bibfield
  {journal} {\bibinfo  {journal} {Journal of Fluid Mechanics}\ }\textbf
  {\bibinfo {volume} {451}},\ \bibinfo {pages} {145 } (\bibinfo {year}
  {2002})}\BibitemShut {NoStop}%
\bibitem [{\citenamefont {Waite}(2011)}]{waite_2011}%
  \BibitemOpen
  \bibfield  {author} {\bibinfo {author} {\bibfnamefont {M.~L.}\ \bibnamefont
  {Waite}},\ }\bibfield  {title} {\bibinfo {title} {Stratified turbulence at
  the buoyancy scale},\ }\href@noop {} {\bibfield  {journal} {\bibinfo
  {journal} {Phys. Fluids}\ }\textbf {\bibinfo {volume} {23}},\ \bibinfo
  {pages} {066602} (\bibinfo {year} {2011})}\BibitemShut {NoStop}%
\bibitem [{\citenamefont {Maffioli}(2017)}]{maffioli_2017}%
  \BibitemOpen
  \bibfield  {author} {\bibinfo {author} {\bibfnamefont {A.}~\bibnamefont
  {Maffioli}},\ }\bibfield  {title} {\bibinfo {title} {Vertical spectra of
  stratified turbulence at large horizontal scales},\ }\href@noop {} {\bibfield
   {journal} {\bibinfo  {journal} {Physical Review Fluids}\ }\textbf {\bibinfo
  {volume} {2}},\ \bibinfo {pages} {104802} (\bibinfo {year}
  {2017})}\BibitemShut {NoStop}%
\bibitem [{\citenamefont {Riley}(2003)}]{riley_2003}%
  \BibitemOpen
  \bibfield  {author} {\bibinfo {author} {\bibfnamefont {J.}~\bibnamefont
  {Riley}},\ }\bibfield  {title} {\bibinfo {title} {Dynamics of turbulence
  strongly influenced by buoyancy},\ }\href@noop {} {\bibfield  {journal}
  {\bibinfo  {journal} {Physics of Fluids}\ }\textbf {\bibinfo {volume} {15}},\
  \bibinfo {pages} {2047} (\bibinfo {year} {2003})}\BibitemShut {NoStop}%
\bibitem [{\citenamefont {van~aartrijk M}\ and\ \citenamefont
  {B}(2008)}]{van_aartrijk_2008}%
  \BibitemOpen
  \bibfield  {author} {\bibinfo {author} {\bibfnamefont {C.~H.}\ \bibnamefont
  {van~aartrijk M}}\ and\ \bibinfo {author} {\bibfnamefont {W.~K.}\
  \bibnamefont {B}},\ }\bibfield  {title} {\bibinfo {title} {Single-particle,
  particle-pair, and multiparticle dispersion of fluid particles in forced
  stably stratified turbulence},\ }\href@noop {} {\bibfield  {journal}
  {\bibinfo  {journal} {Physics of Fluids}\ }\textbf {\bibinfo {volume} {20}},\
  \bibinfo {pages} {025104} (\bibinfo {year} {2008})}\BibitemShut {NoStop}%
\bibitem [{\citenamefont {Sujovolsky}\ and\ \citenamefont
  {Mininni}(2018)}]{Sujo_2018}%
  \BibitemOpen
  \bibfield  {author} {\bibinfo {author} {\bibfnamefont {N.}~\bibnamefont
  {Sujovolsky}}\ and\ \bibinfo {author} {\bibfnamefont {P.}~\bibnamefont
  {Mininni}},\ }\bibfield  {title} {\bibinfo {title} {Vertical dispersion of
  lagrangian tracers in fully developed stably stratified turbulence},\
  }\href@noop {} {\bibfield  {journal} {\bibinfo  {journal} {Physical Review
  Fluids}\ }\textbf {\bibinfo {volume} {4}},\ \bibinfo {pages} {014503}
  (\bibinfo {year} {2018})}\BibitemShut {NoStop}%
\bibitem [{\citenamefont {van aartrijk}\ and\ \citenamefont
  {Clercx}(2010)}]{van_aartrijk_2010}%
  \BibitemOpen
  \bibfield  {author} {\bibinfo {author} {\bibfnamefont {M.}~\bibnamefont {van
  aartrijk}}\ and\ \bibinfo {author} {\bibfnamefont {H.}~\bibnamefont
  {Clercx}},\ }\bibfield  {title} {\bibinfo {title} {Vertical dispersion of
  light inertial particles in stably stratified turbulence: The influence of
  the basset force},\ }\href@noop {} {\bibfield  {journal} {\bibinfo  {journal}
  {Physics of Fluids}\ }\textbf {\bibinfo {volume} {22}},\ \bibinfo {pages} {1}
  (\bibinfo {year} {2010})}\BibitemShut {NoStop}%
\bibitem [{\citenamefont {Sozza}\ \emph {et~al.}(2016)\citenamefont {Sozza},
  \citenamefont {Lillo}, \citenamefont {Musacchio},\ and\ \citenamefont
  {Boffetta}}]{Sozza2016}%
  \BibitemOpen
  \bibfield  {author} {\bibinfo {author} {\bibfnamefont {A.}~\bibnamefont
  {Sozza}}, \bibinfo {author} {\bibfnamefont {F.}~\bibnamefont {Lillo}},
  \bibinfo {author} {\bibfnamefont {S.}~\bibnamefont {Musacchio}},\ and\
  \bibinfo {author} {\bibfnamefont {G.}~\bibnamefont {Boffetta}},\ }\bibfield
  {title} {\bibinfo {title} {Large-scale confinement and small-scale clustering
  of floating particles in stratified turbulence},\ }\href@noop {} {\bibfield
  {journal} {\bibinfo  {journal} {Physical Review Fluids}\ }\textbf {\bibinfo
  {volume} {1}},\ \bibinfo {pages} {1} (\bibinfo {year} {2016})}\BibitemShut
  {NoStop}%
\bibitem [{\citenamefont {Goto}\ and\ \citenamefont
  {Vassilicos}(2008)}]{sweep}%
  \BibitemOpen
  \bibfield  {author} {\bibinfo {author} {\bibfnamefont {S.}~\bibnamefont
  {Goto}}\ and\ \bibinfo {author} {\bibfnamefont {J.}~\bibnamefont
  {Vassilicos}},\ }\bibfield  {title} {\bibinfo {title} {Sweep-stick mechanism
  of heavy particle clustering in fluid turbulence},\ }\href@noop {} {\bibfield
   {journal} {\bibinfo  {journal} {Physical {R}eview {L}etters}\ }\textbf
  {\bibinfo {volume} {100}},\ \bibinfo {pages} {054503} (\bibinfo {year}
  {2008})}\BibitemShut {NoStop}%
\bibitem [{\citenamefont {Obligado}\ \emph {et~al.}(2014)\citenamefont
  {Obligado}, \citenamefont {Teitelbaum}, \citenamefont {Cartellier},
  \citenamefont {Mininni},\ and\ \citenamefont {Bourgoin}}]{obligado}%
  \BibitemOpen
  \bibfield  {author} {\bibinfo {author} {\bibfnamefont {M.}~\bibnamefont
  {Obligado}}, \bibinfo {author} {\bibfnamefont {T.}~\bibnamefont
  {Teitelbaum}}, \bibinfo {author} {\bibfnamefont {A.}~\bibnamefont
  {Cartellier}}, \bibinfo {author} {\bibfnamefont {P.}~\bibnamefont
  {Mininni}},\ and\ \bibinfo {author} {\bibfnamefont {M.}~\bibnamefont
  {Bourgoin}},\ }\bibfield  {title} {\bibinfo {title} {Preferential
  concentration of heavy particles in turbulence},\ }\href@noop {} {\bibfield
  {journal} {\bibinfo  {journal} {Journal of {T}urbulence}\ }\textbf {\bibinfo
  {volume} {15}},\ \bibinfo {pages} {293} (\bibinfo {year} {2014})}\BibitemShut
  {NoStop}%
\bibitem [{\citenamefont {Bragg}\ \emph {et~al.}(2015)\citenamefont {Bragg},
  \citenamefont {Ireland},\ and\ \citenamefont {Collins}}]{Bragg_2015}%
  \BibitemOpen
  \bibfield  {author} {\bibinfo {author} {\bibfnamefont {A.~D.}\ \bibnamefont
  {Bragg}}, \bibinfo {author} {\bibfnamefont {P.~J.}\ \bibnamefont {Ireland}},\
  and\ \bibinfo {author} {\bibfnamefont {L.~R.}\ \bibnamefont {Collins}},\
  }\bibfield  {title} {\bibinfo {title} {Mechanisms for the clustering of
  inertial particles in the inertial range of isotropic turbulence},\
  }\href@noop {} {\bibfield  {journal} {\bibinfo  {journal} {Physical Review
  E}\ }\textbf {\bibinfo {volume} {92}},\ \bibinfo {pages} {023029} (\bibinfo
  {year} {2015})}\BibitemShut {NoStop}%
\bibitem [{\citenamefont {Tom}\ and\ \citenamefont {Bragg}(2019)}]{Tom_2019}%
  \BibitemOpen
  \bibfield  {author} {\bibinfo {author} {\bibfnamefont {J.}~\bibnamefont
  {Tom}}\ and\ \bibinfo {author} {\bibfnamefont {A.~D.}\ \bibnamefont
  {Bragg}},\ }\bibfield  {title} {\bibinfo {title} {Multiscale preferential
  sweeping of particles settling in turbulence},\ }\href@noop {} {\bibfield
  {journal} {\bibinfo  {journal} {Journal of {F}luid {M}echanics}\ }\textbf
  {\bibinfo {volume} {871}},\ \bibinfo {pages} {244} (\bibinfo {year}
  {2019})}\BibitemShut {NoStop}%
\bibitem [{\citenamefont {Homann}\ and\ \citenamefont {Bec}(2009)}]{n1}%
  \BibitemOpen
  \bibfield  {author} {\bibinfo {author} {\bibfnamefont {H.}~\bibnamefont
  {Homann}}\ and\ \bibinfo {author} {\bibfnamefont {J.}~\bibnamefont {Bec}},\
  }\bibfield  {title} {\bibinfo {title} {Finite-size effects in the dynamics of
  neutrally buoyant particles in turbulent flow},\ }\href@noop {} {\bibfield
  {journal} {\bibinfo  {journal} {Journal of {F}luid {M}echanics}\ }\textbf
  {\bibinfo {volume} {651}},\ \bibinfo {pages} {81} (\bibinfo {year}
  {2009})}\BibitemShut {NoStop}%
\bibitem [{\citenamefont {Fiabane}\ \emph {et~al.}(2012)\citenamefont
  {Fiabane}, \citenamefont {Zimmermann}, \citenamefont {Volk}, \citenamefont
  {Pinton},\ and\ \citenamefont {Bourgoin}}]{n2}%
  \BibitemOpen
  \bibfield  {author} {\bibinfo {author} {\bibfnamefont {L.}~\bibnamefont
  {Fiabane}}, \bibinfo {author} {\bibfnamefont {R.}~\bibnamefont {Zimmermann}},
  \bibinfo {author} {\bibfnamefont {R.}~\bibnamefont {Volk}}, \bibinfo {author}
  {\bibfnamefont {J.}~\bibnamefont {Pinton}},\ and\ \bibinfo {author}
  {\bibfnamefont {M.}~\bibnamefont {Bourgoin}},\ }\bibfield  {title} {\bibinfo
  {title} {Clustering of finite-size particles in turbulence},\ }\href@noop {}
  {\bibfield  {journal} {\bibinfo  {journal} {Physical {R}eview {E},
  {S}tatistical, {N}onlinear, and {S}oft {M}atter {P}hysics}\ }\textbf
  {\bibinfo {volume} {86}},\ \bibinfo {pages} {035301} (\bibinfo {year}
  {2012})}\BibitemShut {NoStop}%
\bibitem [{\citenamefont {Angriman}\ \emph {et~al.}(2020)\citenamefont
  {Angriman}, \citenamefont {Mininni},\ and\ \citenamefont {Cobelli}}]{sofi}%
  \BibitemOpen
  \bibfield  {author} {\bibinfo {author} {\bibfnamefont {S.}~\bibnamefont
  {Angriman}}, \bibinfo {author} {\bibfnamefont {P.}~\bibnamefont {Mininni}},\
  and\ \bibinfo {author} {\bibfnamefont {P.}~\bibnamefont {Cobelli}},\
  }\bibfield  {title} {\bibinfo {title} {Velocity and acceleration statistics
  in particle-laden turbulent swirling flows},\ }\href@noop {} {\bibfield
  {journal} {\bibinfo  {journal} {Physical {R}eview {F}luids}\ }\textbf
  {\bibinfo {volume} {5}},\ \bibinfo {pages} {064605} (\bibinfo {year}
  {2020})}\BibitemShut {NoStop}%
\bibitem [{\citenamefont {Moum}(1996)}]{moum_1996}%
  \BibitemOpen
  \bibfield  {author} {\bibinfo {author} {\bibfnamefont {J.~N.}\ \bibnamefont
  {Moum}},\ }\bibfield  {title} {\bibinfo {title} {Energy-containing scales of
  turbulence in the ocean thermocline},\ }\href@noop {} {\bibfield  {journal}
  {\bibinfo  {journal} {Journal of Geophysical Research}\ }\textbf {\bibinfo
  {volume} {101}},\ \bibinfo {pages} {14095} (\bibinfo {year}
  {1996})}\BibitemShut {NoStop}%
\bibitem [{\citenamefont {Clark~di Leoni}\ and\ \citenamefont
  {Mininni}(2015)}]{clark_di_leoni_2015}%
  \BibitemOpen
  \bibfield  {author} {\bibinfo {author} {\bibfnamefont {P.}~\bibnamefont
  {Clark~di Leoni}}\ and\ \bibinfo {author} {\bibfnamefont {P.~D.}\
  \bibnamefont {Mininni}},\ }\bibfield  {title} {\bibinfo {title} {Absorption
  of waves by large-scale winds in stratified turbulence},\ }\href@noop {}
  {\bibfield  {journal} {\bibinfo  {journal} {Phys. Rev. E}\ }\textbf {\bibinfo
  {volume} {91}},\ \bibinfo {pages} {033015} (\bibinfo {year}
  {2015})}\BibitemShut {NoStop}%
\bibitem [{sm()}]{sm}%
  \BibitemOpen
  \href@noop {} {}\bibinfo {note} {For movies showing the early development of
  a mean horizontal wind in the shear layer of the Taylor-Green flow, the
  vertical dispersion of the different particles in the simulation with $N=4$,
  and the horizontal dispersion of particles with $\textrm{St}=1$ in the
  simulation with $N=8$, see the Supplemental Material.}\BibitemShut {Stop}%
\bibitem [{\citenamefont {Mininni}\ \emph {et~al.}(2010)\citenamefont
  {Mininni}, \citenamefont {Rosenberg}, \citenamefont {Reddy},\ and\
  \citenamefont {Pouquet}}]{mininni_hybrid_2011}%
  \BibitemOpen
  \bibfield  {author} {\bibinfo {author} {\bibfnamefont {P.}~\bibnamefont
  {Mininni}}, \bibinfo {author} {\bibfnamefont {D.}~\bibnamefont {Rosenberg}},
  \bibinfo {author} {\bibfnamefont {R.}~\bibnamefont {Reddy}},\ and\ \bibinfo
  {author} {\bibfnamefont {A.}~\bibnamefont {Pouquet}},\ }\bibfield  {title}
  {\bibinfo {title} {A hybrid {MPI}-open{MP} scheme for scalable parallel
  pseudospectral computations for fluid turbulence},\ }\href@noop {} {\bibfield
   {journal} {\bibinfo  {journal} {Parallel Computing}\ }\textbf {\bibinfo
  {volume} {37}},\ \bibinfo {pages} {316} (\bibinfo {year} {2010})}\BibitemShut
  {NoStop}%
\bibitem [{\citenamefont {Yeung}\ and\ \citenamefont
  {Pope}(1988)}]{Yeung_1988}%
  \BibitemOpen
  \bibfield  {author} {\bibinfo {author} {\bibfnamefont {P.}~\bibnamefont
  {Yeung}}\ and\ \bibinfo {author} {\bibfnamefont {S.}~\bibnamefont {Pope}},\
  }\bibfield  {title} {\bibinfo {title} {An algorithm for tracking fluid
  particles in numerical simulations of homogeneous turbulence},\ }\href@noop
  {} {\bibfield  {journal} {\bibinfo  {journal} {Journal of Computational
  Physics}\ }\textbf {\bibinfo {volume} {79}},\ \bibinfo {pages} {373}
  (\bibinfo {year} {1988})}\BibitemShut {NoStop}%
\bibitem [{\citenamefont {Donzis}\ and\ \citenamefont
  {Yeung}(2010)}]{Donzis_2010}%
  \BibitemOpen
  \bibfield  {author} {\bibinfo {author} {\bibfnamefont {D.}~\bibnamefont
  {Donzis}}\ and\ \bibinfo {author} {\bibfnamefont {P.}~\bibnamefont {Yeung}},\
  }\bibfield  {title} {\bibinfo {title} {Resolution effects and scaling in
  numerical simulations of passive scalar mixing in turbulence},\ }\href@noop
  {} {\bibfield  {journal} {\bibinfo  {journal} {Physica D: Nonlinear
  Phenomena}\ }\textbf {\bibinfo {volume} {239}},\ \bibinfo {pages} {1278}
  (\bibinfo {year} {2010})}\BibitemShut {NoStop}%
\bibitem [{\citenamefont {Wan}\ \emph {et~al.}(2010)\citenamefont {Wan},
  \citenamefont {Oughton}, \citenamefont {Servidio},\ and\ \citenamefont
  {Matthaeus}}]{Wan_2010}%
  \BibitemOpen
  \bibfield  {author} {\bibinfo {author} {\bibfnamefont {M.}~\bibnamefont
  {Wan}}, \bibinfo {author} {\bibfnamefont {S.}~\bibnamefont {Oughton}},
  \bibinfo {author} {\bibfnamefont {S.}~\bibnamefont {Servidio}},\ and\
  \bibinfo {author} {\bibfnamefont {W.~H.}\ \bibnamefont {Matthaeus}},\
  }\bibfield  {title} {\bibinfo {title} {On the accuracy of simulations of
  turbulence},\ }\href@noop {} {\bibfield  {journal} {\bibinfo  {journal}
  {Physics of Plasmas}\ }\textbf {\bibinfo {volume} {17}},\ \bibinfo {pages}
  {082308} (\bibinfo {year} {2010})}\BibitemShut {NoStop}%
\bibitem [{\citenamefont {D'Asaro}\ \emph {et~al.}(2007)\citenamefont
  {D'Asaro}, \citenamefont {Lien},\ and\ \citenamefont
  {Henyey}}]{D'Asaro_2007}%
  \BibitemOpen
  \bibfield  {author} {\bibinfo {author} {\bibfnamefont {E.}~\bibnamefont
  {D'Asaro}}, \bibinfo {author} {\bibfnamefont {R.~C.}\ \bibnamefont {Lien}},\
  and\ \bibinfo {author} {\bibfnamefont {F.}~\bibnamefont {Henyey}},\
  }\bibfield  {title} {\bibinfo {title} {High-frequency internal waves on the
  oregon continental shelf},\ }\href@noop {} {\bibfield  {journal} {\bibinfo
  {journal} {Journal of Physical Oceanography}\ }\textbf {\bibinfo {volume}
  {37}},\ \bibinfo {pages} {1} (\bibinfo {year} {2007})}\BibitemShut {NoStop}%
\bibitem [{\citenamefont {Sujovolsky}\ and\ \citenamefont
  {Mininni}(2019)}]{Sujo_lag_2019}%
  \BibitemOpen
  \bibfield  {author} {\bibinfo {author} {\bibfnamefont {N.~E.}\ \bibnamefont
  {Sujovolsky}}\ and\ \bibinfo {author} {\bibfnamefont {P.~D.}\ \bibnamefont
  {Mininni}},\ }\bibfield  {title} {\bibinfo {title} {Vertical dispersion of
  lagrangian tracers in fully developed stably stratified turbulence},\
  }\href@noop {} {\bibfield  {journal} {\bibinfo  {journal} {Phys. Rev.
  Fluids}\ }\textbf {\bibinfo {volume} {4}},\ \bibinfo {pages} {014503}
  (\bibinfo {year} {2019})}\BibitemShut {NoStop}%
\bibitem [{\citenamefont {Nicolleau}\ and\ \citenamefont
  {Vassilicos}(2000)}]{nicolleau_2000}%
  \BibitemOpen
  \bibfield  {author} {\bibinfo {author} {\bibfnamefont {F.}~\bibnamefont
  {Nicolleau}}\ and\ \bibinfo {author} {\bibfnamefont {J.~C.}\ \bibnamefont
  {Vassilicos}},\ }\bibfield  {title} {\bibinfo {title} {Turbulent diffusion in
  stably stratified non-decaying turbulence},\ }\href@noop {} {\bibfield
  {journal} {\bibinfo  {journal} {Journal of Fluid Mechanics}\ }\textbf
  {\bibinfo {volume} {410}},\ \bibinfo {pages} {123} (\bibinfo {year}
  {2000})}\BibitemShut {NoStop}%
\bibitem [{\citenamefont {Feraco}\ \emph {et~al.}(2018)\citenamefont {Feraco},
  \citenamefont {Marino}, \citenamefont {Pumir}, \citenamefont {Primavera},
  \citenamefont {Mininni}, \citenamefont {Pouquet},\ and\ \citenamefont
  {Rosenberg}}]{Feraco_2018}%
  \BibitemOpen
  \bibfield  {author} {\bibinfo {author} {\bibfnamefont {F.}~\bibnamefont
  {Feraco}}, \bibinfo {author} {\bibfnamefont {R.}~\bibnamefont {Marino}},
  \bibinfo {author} {\bibfnamefont {A.}~\bibnamefont {Pumir}}, \bibinfo
  {author} {\bibfnamefont {L.}~\bibnamefont {Primavera}}, \bibinfo {author}
  {\bibfnamefont {P.~D.}\ \bibnamefont {Mininni}}, \bibinfo {author}
  {\bibfnamefont {A.}~\bibnamefont {Pouquet}},\ and\ \bibinfo {author}
  {\bibfnamefont {D.}~\bibnamefont {Rosenberg}},\ }\bibfield  {title} {\bibinfo
  {title} {Vertical drafts and mixing in stratified turbulence: Sharp
  transition with froude number},\ }\href@noop {} {\bibfield  {journal}
  {\bibinfo  {journal} {{EPL} (Europhysics Letters)}\ }\textbf {\bibinfo
  {volume} {123}},\ \bibinfo {pages} {44002} (\bibinfo {year}
  {2018})}\BibitemShut {NoStop}%
\bibitem [{\citenamefont {Sozza}\ \emph {et~al.}(2018)\citenamefont {Sozza},
  \citenamefont {De~Lillo},\ and\ \citenamefont {Boffetta}}]{sozza_2018}%
  \BibitemOpen
  \bibfield  {author} {\bibinfo {author} {\bibfnamefont {A.}~\bibnamefont
  {Sozza}}, \bibinfo {author} {\bibfnamefont {F.}~\bibnamefont {De~Lillo}},\
  and\ \bibinfo {author} {\bibfnamefont {G.}~\bibnamefont {Boffetta}},\
  }\bibfield  {title} {\bibinfo {title} {Inertial floaters in stratified
  turbulence},\ }\href@noop {} {\bibfield  {journal} {\bibinfo  {journal} {EPL
  (Europhysics Letters)}\ }\textbf {\bibinfo {volume} {121}},\ \bibinfo {pages}
  {14002} (\bibinfo {year} {2018})}\BibitemShut {NoStop}%
\bibitem [{\citenamefont {Pugliese}\ and\ \citenamefont
  {Dmitruk}(2022)}]{Pugliese_2022}%
  \BibitemOpen
  \bibfield  {author} {\bibinfo {author} {\bibfnamefont {F.}~\bibnamefont
  {Pugliese}}\ and\ \bibinfo {author} {\bibfnamefont {P.}~\bibnamefont
  {Dmitruk}},\ }\bibfield  {title} {\bibinfo {title} {Test particle
  energization of heavy ions in magnetohydrodynamic turbulence},\ }\href@noop
  {} {\bibfield  {journal} {\bibinfo  {journal} {The Astrophysical Journal}\
  }\textbf {\bibinfo {volume} {929}},\ \bibinfo {pages} {4} (\bibinfo {year}
  {2022})}\BibitemShut {NoStop}%
\bibitem [{\citenamefont {Monchaux}\ \emph {et~al.}(2010)\citenamefont
  {Monchaux}, \citenamefont {Bourgoin},\ and\ \citenamefont
  {Cartellier}}]{vor15}%
  \BibitemOpen
  \bibfield  {author} {\bibinfo {author} {\bibfnamefont {R.}~\bibnamefont
  {Monchaux}}, \bibinfo {author} {\bibfnamefont {M.}~\bibnamefont {Bourgoin}},\
  and\ \bibinfo {author} {\bibfnamefont {A.}~\bibnamefont {Cartellier}},\
  }\bibfield  {title} {\bibinfo {title} {Preferential concentration of heavy
  particles: A {V}oronoï analysis},\ }\href@noop {} {\bibfield  {journal}
  {\bibinfo  {journal} {Physics of {F}luids}\ }\textbf {\bibinfo {volume}
  {22}},\ \bibinfo {pages} {103304} (\bibinfo {year} {2010})}\BibitemShut
  {NoStop}%
\bibitem [{\citenamefont {Monchaux}\ \emph {et~al.}(2012)\citenamefont
  {Monchaux}, \citenamefont {Bourgoin},\ and\ \citenamefont
  {Cartellier}}]{vor16}%
  \BibitemOpen
  \bibfield  {author} {\bibinfo {author} {\bibfnamefont {R.}~\bibnamefont
  {Monchaux}}, \bibinfo {author} {\bibfnamefont {M.}~\bibnamefont {Bourgoin}},\
  and\ \bibinfo {author} {\bibfnamefont {A.}~\bibnamefont {Cartellier}},\
  }\bibfield  {title} {\bibinfo {title} {Analyzing preferential concentration
  and clustering of inertial particles in turbulence},\ }\href@noop {}
  {\bibfield  {journal} {\bibinfo  {journal} {International {J}ournal of
  {M}ultiphase {F}low}\ }\textbf {\bibinfo {volume} {40}},\ \bibinfo {pages}
  {1} (\bibinfo {year} {2012})}\BibitemShut {NoStop}%
\bibitem [{\citenamefont {Sumbekova}\ \emph {et~al.}(2017)\citenamefont
  {Sumbekova}, \citenamefont {Cartellier}, \citenamefont {Aliseda},\ and\
  \citenamefont {Bourgoin}}]{Sumbekova_2017}%
  \BibitemOpen
  \bibfield  {author} {\bibinfo {author} {\bibfnamefont {S.}~\bibnamefont
  {Sumbekova}}, \bibinfo {author} {\bibfnamefont {A.}~\bibnamefont
  {Cartellier}}, \bibinfo {author} {\bibfnamefont {A.}~\bibnamefont
  {Aliseda}},\ and\ \bibinfo {author} {\bibfnamefont {M.}~\bibnamefont
  {Bourgoin}},\ }\bibfield  {title} {\bibinfo {title} {Preferential
  concentration of inertial sub-{K}olmogorov particles: The roles of mass
  loading of particles, stokes numbers, and reynolds numbers},\ }\href@noop {}
  {\bibfield  {journal} {\bibinfo  {journal} {Physical Review Fluids}\ }\textbf
  {\bibinfo {volume} {2}},\ \bibinfo {pages} {024302} (\bibinfo {year}
  {2017})}\BibitemShut {NoStop}%
\bibitem [{\citenamefont {Obligado}\ \emph {et~al.}(2020)\citenamefont
  {Obligado}, \citenamefont {Cartellier}, \citenamefont {Aliseda},
  \citenamefont {Calmant},\ and\ \citenamefont {de~Palma}}]{Obligado_2020}%
  \BibitemOpen
  \bibfield  {author} {\bibinfo {author} {\bibfnamefont {M.}~\bibnamefont
  {Obligado}}, \bibinfo {author} {\bibfnamefont {A.}~\bibnamefont
  {Cartellier}}, \bibinfo {author} {\bibfnamefont {A.}~\bibnamefont {Aliseda}},
  \bibinfo {author} {\bibfnamefont {T.}~\bibnamefont {Calmant}},\ and\ \bibinfo
  {author} {\bibfnamefont {N.}~\bibnamefont {de~Palma}},\ }\bibfield  {title}
  {\bibinfo {title} {Study on preferential concentration of inertial particles
  in homogeneous isotropic turbulence via big-data techniques},\ }\href@noop {}
  {\bibfield  {journal} {\bibinfo  {journal} {Physical Review Fluids}\ }\textbf
  {\bibinfo {volume} {5}},\ \bibinfo {pages} {024303} (\bibinfo {year}
  {2020})}\BibitemShut {NoStop}%
\bibitem [{\citenamefont {van Aartrijk}\ and\ \citenamefont
  {Clercx}(2008)}]{Aartrijk_prefential_2008}%
  \BibitemOpen
  \bibfield  {author} {\bibinfo {author} {\bibfnamefont {M.}~\bibnamefont {van
  Aartrijk}}\ and\ \bibinfo {author} {\bibfnamefont {H.~J.~H.}\ \bibnamefont
  {Clercx}},\ }\bibfield  {title} {\bibinfo {title} {Preferential concentration
  of heavy particles in stably stratified turbulence},\ }\href@noop {}
  {\bibfield  {journal} {\bibinfo  {journal} {Phys. Rev. Lett.}\ }\textbf
  {\bibinfo {volume} {100}},\ \bibinfo {pages} {254501} (\bibinfo {year}
  {2008})}\BibitemShut {NoStop}%
\bibitem [{\citenamefont {Tanemura}(2003)}]{Tanemura_2003}%
  \BibitemOpen
  \bibfield  {author} {\bibinfo {author} {\bibfnamefont {M.}~\bibnamefont
  {Tanemura}},\ }\bibfield  {title} {\bibinfo {title} {Statistical
  distributions of poisson voronoi cells in two and three dimensions},\
  }\href@noop {} {\bibfield  {journal} {\bibinfo  {journal} {Forma}\ }\textbf
  {\bibinfo {volume} {18}},\ \bibinfo {pages} {221} (\bibinfo {year}
  {2003})}\BibitemShut {NoStop}%
\bibitem [{\citenamefont {Uhlmann}(2020)}]{Uhlmann_2020}%
  \BibitemOpen
  \bibfield  {author} {\bibinfo {author} {\bibfnamefont {M.}~\bibnamefont
  {Uhlmann}},\ }\bibfield  {title} {\bibinfo {title} {Vorono{\"\i} tessellation
  analysis of sets of randomly placed finite-size spheres},\ }\href@noop {}
  {\bibfield  {journal} {\bibinfo  {journal} {Physica A}\ }\textbf {\bibinfo
  {volume} {555}},\ \bibinfo {pages} {124618} (\bibinfo {year}
  {2020})}\BibitemShut {NoStop}%
\bibitem [{\citenamefont {Reartes}\ and\ \citenamefont
  {Mininni}(2021)}]{Reartes_2021}%
  \BibitemOpen
  \bibfield  {author} {\bibinfo {author} {\bibfnamefont {C.}~\bibnamefont
  {Reartes}}\ and\ \bibinfo {author} {\bibfnamefont {P.}~\bibnamefont
  {Mininni}},\ }\bibfield  {title} {\bibinfo {title} {Settling and clustering
  of particles of moderate mass density in turbulence},\ }\href@noop {}
  {\bibfield  {journal} {\bibinfo  {journal} {Phys. Rev. Fluids}\ }\textbf
  {\bibinfo {volume} {6}},\ \bibinfo {pages} {114304} (\bibinfo {year}
  {2021})}\BibitemShut {NoStop}%
\bibitem [{\citenamefont {De~Pietro}\ \emph {et~al.}(2014)\citenamefont
  {De~Pietro}, \citenamefont {van Hinsberg}, \citenamefont {Biferale},
  \citenamefont {Clercx}, \citenamefont {Perlekar},\ and\ \citenamefont
  {Toschi}}]{n5}%
  \BibitemOpen
  \bibfield  {author} {\bibinfo {author} {\bibfnamefont {M.}~\bibnamefont
  {De~Pietro}}, \bibinfo {author} {\bibfnamefont {M.}~\bibnamefont {van
  Hinsberg}}, \bibinfo {author} {\bibfnamefont {L.}~\bibnamefont {Biferale}},
  \bibinfo {author} {\bibfnamefont {H.}~\bibnamefont {Clercx}}, \bibinfo
  {author} {\bibfnamefont {P.}~\bibnamefont {Perlekar}},\ and\ \bibinfo
  {author} {\bibfnamefont {F.}~\bibnamefont {Toschi}},\ }\bibfield  {title}
  {\bibinfo {title} {Clustering of vertically constrained passive particles in
  homogeneous, isotropic turbulence},\ }\href@noop {} {\bibfield  {journal}
  {\bibinfo  {journal} {Physical {R}eview {E}}\ }\textbf {\bibinfo {volume}
  {91}},\ \bibinfo {pages} {053002} (\bibinfo {year} {2014})}\BibitemShut
  {NoStop}%
\bibitem [{\citenamefont {Shaw}(2003)}]{Shaw_2003}%
  \BibitemOpen
  \bibfield  {author} {\bibinfo {author} {\bibfnamefont {R.~A.}\ \bibnamefont
  {Shaw}},\ }\bibfield  {title} {\bibinfo {title} {Particle-turbulence
  interactions in atmospheric clouds},\ }\href@noop {} {\bibfield  {journal}
  {\bibinfo  {journal} {Annual Review of Fluid Mechanics}\ }\textbf {\bibinfo
  {volume} {35}},\ \bibinfo {pages} {183} (\bibinfo {year} {2003})}\BibitemShut
  {NoStop}%
\bibitem [{\citenamefont {Squires}\ and\ \citenamefont
  {Yamazaki}(1995)}]{Squires_1995}%
  \BibitemOpen
  \bibfield  {author} {\bibinfo {author} {\bibfnamefont {K.~D.}\ \bibnamefont
  {Squires}}\ and\ \bibinfo {author} {\bibfnamefont {H.}~\bibnamefont
  {Yamazaki}},\ }\bibfield  {title} {\bibinfo {title} {Preferential
  concentration of marine particles in isotropic turbulence},\ }\href@noop {}
  {\bibfield  {journal} {\bibinfo  {journal} {Deep Sea Research Part I:
  Oceanographic Research Papers}\ }\textbf {\bibinfo {volume} {42}},\ \bibinfo
  {pages} {1989} (\bibinfo {year} {1995})}\BibitemShut {NoStop}%
\bibitem [{\citenamefont {Martin}(2003)}]{Martin_2003}%
  \BibitemOpen
  \bibfield  {author} {\bibinfo {author} {\bibfnamefont {A.}~\bibnamefont
  {Martin}},\ }\bibfield  {title} {\bibinfo {title} {Phytoplankton patchiness:
  the role of lateral stirring and mixing},\ }\href@noop {} {\bibfield
  {journal} {\bibinfo  {journal} {Progress in Oceanography}\ }\textbf {\bibinfo
  {volume} {57}},\ \bibinfo {pages} {125} (\bibinfo {year} {2003})}\BibitemShut
  {NoStop}%
\bibitem [{\citenamefont {Durham}\ \emph {et~al.}(2013)\citenamefont {Durham},
  \citenamefont {Climent}, \citenamefont {Barry}, \citenamefont {Lillo},
  \citenamefont {Boffetta}, \citenamefont {Cencini},\ and\ \citenamefont
  {Stocker}}]{Durham_2013}%
  \BibitemOpen
  \bibfield  {author} {\bibinfo {author} {\bibfnamefont {W.~M.}\ \bibnamefont
  {Durham}}, \bibinfo {author} {\bibfnamefont {E.}~\bibnamefont {Climent}},
  \bibinfo {author} {\bibfnamefont {M.}~\bibnamefont {Barry}}, \bibinfo
  {author} {\bibfnamefont {F.~D.}\ \bibnamefont {Lillo}}, \bibinfo {author}
  {\bibfnamefont {G.}~\bibnamefont {Boffetta}}, \bibinfo {author}
  {\bibfnamefont {M.}~\bibnamefont {Cencini}},\ and\ \bibinfo {author}
  {\bibfnamefont {R.}~\bibnamefont {Stocker}},\ }\bibfield  {title} {\bibinfo
  {title} {Turbulence drives microscale patches of motile phytoplankton},\
  }\href@noop {} {\bibfield  {journal} {\bibinfo  {journal} {Nature
  Communications}\ }\textbf {\bibinfo {volume} {4}},\ \bibinfo {pages} {2148}
  (\bibinfo {year} {2013})}\BibitemShut {NoStop}%
\bibitem [{\citenamefont {Del~Grosso}\ \emph {et~al.}(2019)\citenamefont
  {Del~Grosso}, \citenamefont {Cappelletti}, \citenamefont {Sujovolsky},
  \citenamefont {Mininni},\ and\ \citenamefont {Cobelli}}]{Del_Grosso_2019}%
  \BibitemOpen
  \bibfield  {author} {\bibinfo {author} {\bibfnamefont {N.~F.}\ \bibnamefont
  {Del~Grosso}}, \bibinfo {author} {\bibfnamefont {L.~M.}\ \bibnamefont
  {Cappelletti}}, \bibinfo {author} {\bibfnamefont {N.~E.}\ \bibnamefont
  {Sujovolsky}}, \bibinfo {author} {\bibfnamefont {P.~D.}\ \bibnamefont
  {Mininni}},\ and\ \bibinfo {author} {\bibfnamefont {P.~J.}\ \bibnamefont
  {Cobelli}},\ }\bibfield  {title} {\bibinfo {title} {Statistics of single and
  multiple floaters in experiments of surface wave turbulence},\ }\href@noop {}
  {\bibfield  {journal} {\bibinfo  {journal} {Physical Review Fluids}\ }\textbf
  {\bibinfo {volume} {4}},\ \bibinfo {pages} {074805} (\bibinfo {year}
  {2019})}\BibitemShut {NoStop}%
\end{thebibliography}%

\end{document}